\documentclass{article}
\usepackage{graphicx}\usepackage{epsfig}
\usepackage{amsmath}\usepackage{amssymb}\usepackage{amsfonts}
\usepackage{bm}
\usepackage{exscale}
\usepackage{dcolumn}    
\usepackage{color}
\usepackage{soul}
\usepackage{ulem}
\usepackage{sidecap}
\usepackage{mathrsfs} 
\usepackage{texnames}
\usepackage[T1]{fontenc}

\sidecaptionvpos{figure}{t}
\begin{document}

\title{\textbf{The symmetry energy of the nuclear EoS: \\a study\,of\,collective\,motion and\,low-energy\\reaction\,dynamics\,in\,semiclassical\,approaches}}

\author{S. Burrello\,$^{1}$, M. Colonna\,$^{1}$, H. Zheng\,$^{1,2}$}

\date{}

\maketitle

\begin{center}
{$^{1}$ Laboratori Nazionali del Sud, INFN, 95123 Catania, Italy \\}
{$^{2}$School  of  Physics  and  Information  Technology, Shaanxi  Normal  University,  Xi'an  710119,  China}
\vspace*{0.5cm}
\setlength \parindent{0 cm}
\end{center}

\begin{abstract}
In the framework of mean-field based transport approaches, we discuss recent results concerning 
collective motion and low-energy
heavy ion reactions involving neutron-rich systems.
We focus on aspects which are particularly sensitive to  
the isovector terms of the nuclear effective interaction and the corresponding symmetry energy. 
As far as collective excitations are concerned, we discuss the mixed nature of dipole oscillations in neutron-rich 
systems. On the other hand, for reactions close to the Coulomb barrier, we 
investigate the structure of pre-equilibrium 
collective dipole oscillations, focusing on their sensitivity  to the 
symmetry energy behavior below normal density. 
Nucleon emission is also considered within the same context. 
The possible impact of
other relevant terms of the nuclear effective interaction on these mechanisms 
is also examined. 
From this analysis we expect to put further constraints on the nuclear Equation of State, 
of crucial importance also in the astrophysical context.
\end{abstract}

\section{Introduction}


Collective patterns exhibited by complex systems can bear important information on 
{relevant properties of the particle interaction}. 
{In nuclei, the investigation of the giant resonances, whose collective nature is well established, is therefore of primary importance~\cite{Har2001}. A prominent example in this context is the giant dipole resonance (GDR), which can be described 
in terms of {protons and neutrons 
oscillating as a whole against each other.}

Stimulated by the advent of new radioactive beam facilities, a large amount of research has been devoted in recent years to the features of unstable nuclei and their collective multipole response. In the case of nuclei with some neutron excess, a strong fragmentation 
of strength has been observed in the isovector dipole response, mainly located 
{at lower energy with respect to the GDR}~\cite{harPRL2000,harPRC2002,konPRC2012,adrPRL2005,kliPRC2007,carPRC2010,wiePPNP2011,tamPRL2011,ni3,sn2}. These low-lying  excitations, which are referred in literature as Pygmy Dipole Resonance (PDR), have been the object of intense discussion~\cite{aumPS2013,savPPNP2013,paaRPP2007,lanPRC2011,Crespi2014,Edo1,Endre,Repko2013,Papa,Knapp1,Knapp2,Auer}, proving to be, likewise the GDR, an important probe of crucial information of the nuclear effective interaction, especially concerning its isovector component and the corresponding contribution to the Equation of State (EoS)~\cite{barPR2005,li08,giuliani14}, namely the symmetry energy. 

Collective oscillations of neutrons against protons might occur also in low-energy reactions involving charge-asymmetric systems, at least during the pre-equilibrium stage. If the N/Z ratios of the reaction partners are appreciably different, then neutron and proton centers of mass of the involved composite system do not coincide in the early phase of the fusion path and charge equilibration mechanisms take place. As a result, together with the incoherent exchange of nucleons between the reacting ions, a dynamical dipole (DD) mode, also kwnown as pre-equilibrium GDR~\cite{baranPRL2001,simenelPRL2001,Umar07,
wuPRC2010,flibottePRL1996,Papa05,pierrPRC2009,Giaz14} is observed, along the symmetry axis of the dinuclear system. 

Since the transient composite system might experience large prolate deformation with respect to the equilibium configuration of the final compound nucleus, the corresponding pre-equilibrium radiation carries out fundamental information about the density distribution and the shape of the di-nuclear complex. It is worth noting that this mechanism may also provide 
a cooling effect, which could favour superheavy element formation~\cite{simenelPRC2007,parasPRC2016}. 

Apart from the strong influence of different parameters, such as mass and charge asymmetry, collision centrality and energy~\cite{baranPRL2001,pierrPRC2009,tson2001}, collective oscillations which characterize the DD turn out to be mainly ruled by the isovector channel of the nuclear effective interaction, which 
yields once again the restoring force. 
{However, within the selected beam energy (around 10 MeV/A), where the DD mechanism 
is better evidenced, other pre-equilibrium effects, such as nucleon and light particle emission, 
can occur, leading to a reduction of the initial charge asymmetry of the colliding nuclei and 
contributing to cool down the system. Likewise the DD mechanism, also the N/Z ratio of the pre-equilibrium nucleon emission has been proposed as a 
probe of the symmetry energy behavior below normal density~\cite{pierrPRC2009,Giaz14,baranPRC2009}.} } 

In this article we review recent studies devoted to the investigation, within a semi-classical
transport approach,  of collective excitations  in isolated nuclei and of pre-equilibrium 
{effects, such as dipole radiation and nucleon emission, occurring  
in nuclear reactions at low beam energy}~\cite{zhePRC2016,zhePLB}.
The nuclear effective interaction is described by Skyrme-like parameterizations, 
which are mainly tuned on the features
of selected nuclei, especially in spin-isospin channels~\cite{coll4}.
We will explore the sensitivity of the mechanisms considered to specific properties of
the effective interaction and, in the case of nuclear reactions, also to the strength of two-body (n-n) collision
cross section.
In particular, from our combined analysis, we aim at getting a consistent picture of the impact of 
the density dependence of the symmetry energy on dipole excitations in neutron-rich systems 
and on the features of pre-equilibrium DD oscillations, together with nucleon emission. 
{We stress the general interest of this study,
considering the leading role played by the symmetry energy in 
nuclear structure problems (the neutron skin thickness, for instance)~\cite{piek12, GaiPRC2011, GaiPRC2012}
and its impact in the astrophysical context~\cite{stePR2005,burrelloPRC2015}. }    

\section{Theoretical framework}

{We adopt here the same theoretical and numerical treatments illustrated
on Refs.~\cite{zhePRC2016,zhePLB}, namely calculations are based on  
the semi-classical Boltzmann-Nordheim-Vlasov (BNV) model~\cite{Bon94,Guar96}.}

Within such a framework, the evolution of the system is investigated by solving the two dynamical coupled equations~\cite{barPR2005}:
\begin{equation}
\frac{\partial f_q({\bf r},{\bf p},t)}{\partial t}+\frac{\partial \epsilon_q}{\partial {\bf p}}\frac{\partial f_q ({\bf r},{\bf p},t)}{\partial {\bf r}}-
\frac{\partial \epsilon_q}{\partial {\bf r}}\frac{\partial f_q({\bf r},{\bf p},t)}{\partial {\bf p}}= I_{coll}[f_n,f_p] ,
\label{vlasov}
\end{equation}
where $f_q$ and  $\epsilon_q$, with q = n, p, are the distribution functions and the single particle energies of neutrons
and protons, respectively. In the spirit of the density functional theory, the single particle energy, which includes
the mean-field potential, can be derived 
from an energy density functional, $\mathscr{E}$~\cite{Lar98}. The latter
quantity, in the case of Skyrme-like interactions, is written as~\cite{radutaEJPA2014}: 
\begin{align}
\mathscr{E} &= \frac{\hbar^2}{2 m}\tau + C_0\rho^2 + D_0\rho_{3}^2 + C_3\rho^{\alpha + 2} + D_3\rho^{\alpha}\rho_{3}^2 ~+ \nonumber \\ 
& C_{eff}\rho\tau + D_{eff}\rho_{3}\tau_{3} + C_{surf}(\bigtriangledown\rho)^2 + D_{surf}(\bigtriangledown\rho_3)^2,
\label{eq:rhoE}
\end{align}
where ($\rho=\rho_n+\rho_p, \rho_{3}=\rho_n-\rho_p$) 
and ($\tau=\tau_{n}+\tau_{p},  \tau_{3}=\tau_{n}-\tau_{p}$)
denote isoscalar and isovector
density and kinetic energy densities, respectively, and the standard Skyrme parameters have been properly
combined into the coefficients  $C_{..}$, $D_{..}$.
In the calculations, the Coulomb contribution is also 
included~\cite{zhePRC2016}.
The
effect of the residual two-body correlations is taken into account in the collision integral, $I_{coll}[f_n,f_p]$, 
employing the isospin, energy and angular dependent free nucleon-nucleon cross section.
The test-particle (t.p.) method~\cite{wong} is adopted to integrate Eq.~(\ref{vlasov}). However, the finite number of t.p.
considered requires to set a maximum cutoff of 50 mb for the n-n 
cross section~\cite{Bar02,Akira07}, to quench spurious collisions
that may originate from an inaccurate evaluation of Pauli blocking 
effects. 

{The model illustrated here is able to describe nuclear dynamics at low 
beam energies, from fusion to quasi-fission and deep-inelastic processes~\cite{pierrPRC2009,Rizzo14}.
Moreover, the features of zero-sound excitations are well reproduced, 
both in nuclear matter and finite nuclei~\cite{barPR2005, zhePRC2016, urbPRC2012}, though quantum effects, such as shell effects, cannot be accounted for. }

Among the different channels of the effective interaction, we are mainly interested in the isovector terms. Thus, we introduce the
definition of  
the symmetry energy per nucleon, $E_{sym}/ A = C(\rho) I^2$, 
where $I = \rho_3/\rho$ is the asymmetry parameter. 
The coefficient $C(\rho)$ can be expressed in terms of the Skyrme coefficients:
\begin{equation}
C(\rho) = \frac{\varepsilon_F}{3} + D_0\rho + D_3\rho^{\alpha+1} ~+ 
\frac{2m}{\hbar^2}\left(\frac{C_{eff}}{3} + D_{eff}\right)\varepsilon_F\rho,
\end{equation}
where $\varepsilon_F$ denotes the Fermi energy and $m$ is the nucleon mass.   

In our calculations we will employ the recently introduced SAMi-J Skyrme effective interactions. 
The details of the SAMi fitting protocol and the derivation of the corresponding parameters
can be found in Ref.~\cite{coll4}.
As a key feature, the SAMi-J family has been produced to allow for different values of the symmetry
energy at normal density, $J = C(\rho_0)$, from 27 to 35 MeV, but keeping the same optimal values
of the main isoscalar nuclear matter properties and of the main features of selected finite nuclei. 
In this way, these interactions mainly differ in the isovector channel and are thus well suited 
to explore the impact of isovector terms on a given observable.  

\begin{figure}
\begin{center}
\includegraphics*[scale=0.36]{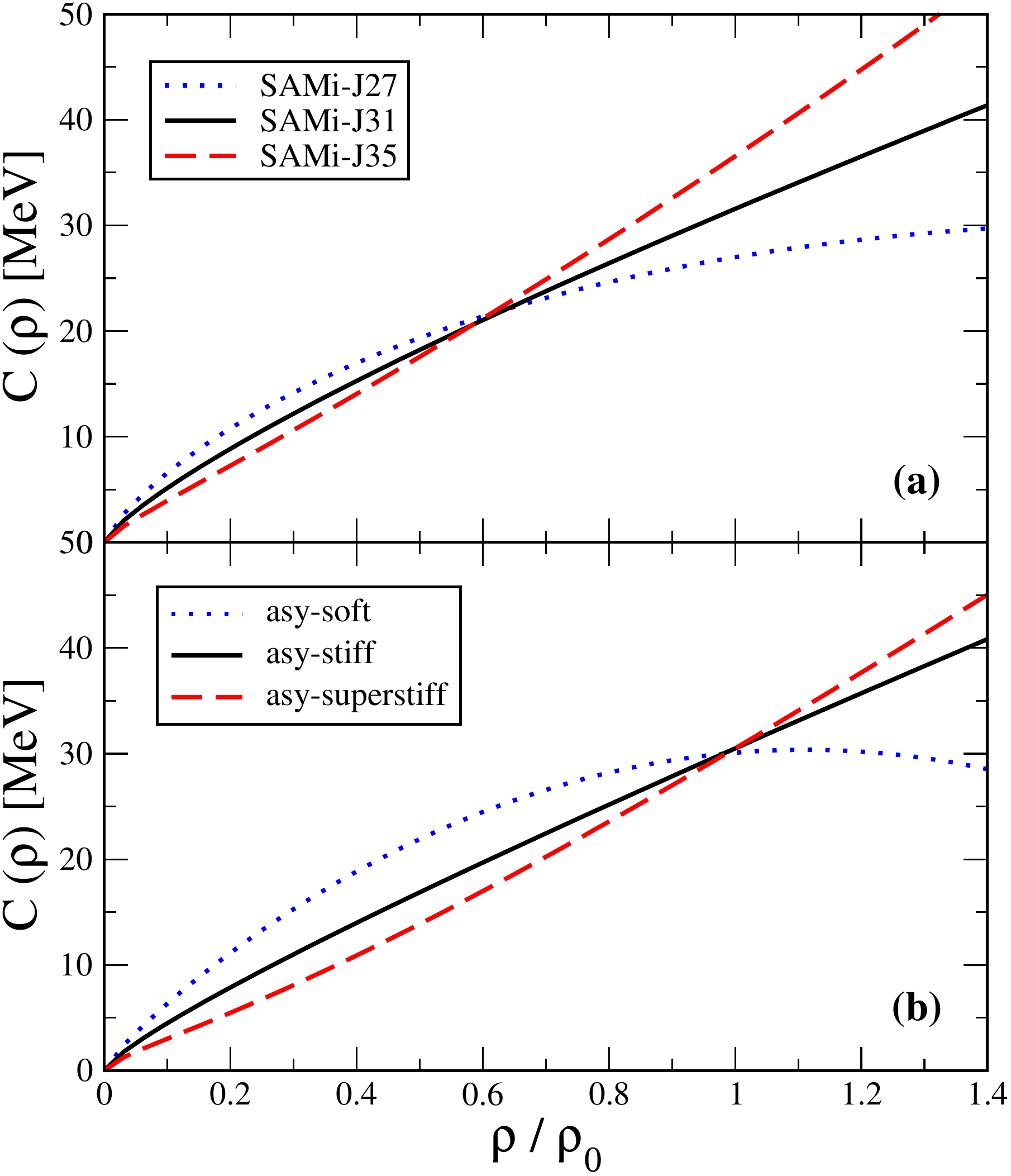}
\end{center}
\caption{(Color online) 
{Density dependence of the symmetry energy for the EoS with (upper panel) or without (lower panel) momentum dependence ($\rho_0 = 0.16$ fm$^{-3}$}). Readapted from~\cite{zhePLB}.}
\label{eossym}
\end{figure}
It is worth noting here that, by construction, the Skyrme mean-field potential $U_q$ is associated with
a quadratic dependence on the momentum. This behavior is a good approximation for low momenta,
such as in the situation explored in our study~\cite{bao_PLB}. Actually, for the SAMi-J interactions, a rather flat momentum dependence  
is observed for the symmetry potential, according to the small 
splitting, $m^*_n - m^*_p = 0.023~mI$, 
between neutron and proton effective masses. Moreover, 
an effective isoscalar mass $m^*(I = 0) = 0.67~m$ MeV is predicted 
by these interactions.

{In the following, we will consider}
three SAMi-J parametrizations: SAMi-J27, SAMi-J31 and SAMi-J35~\cite{coll4}. 
Since the fitting procedure involves the properties of 
finite nuclei, 
the coefficient  $C(\rho)$ gets the same value, i.e.  $C(\rho_c) \approx 22$ 
MeV at the density $\rho_c \approx 0.6\rho_0$, that approximately
represents the average density of nuclei of intermediate mass.
Consequently, each parametrization is characterized by a different symmentry
energy value, J, at normal density, as indicated in the corresponding 
interaction name.    


The values of the slope parameter $\displaystyle L = 3 \left. \rho_0 \frac{d C(\rho)}{d \rho} \right\vert_{\rho=\rho_0}$ are reported in Table 1. The corresponding density dependence of $C(\rho)$ is displayed in Fig.~\ref{eossym}(a).

We will also adopt momentum independent Skyrme interactions ($C_{eff} = D_{eff} = 0$, 
$m^* = m$), characterized by an incompressibility modulus $K = 200$ MeV~\cite{coll2} and widely employed in the literature~\cite{baranPRC2009,zhePRC2016,coll2}.  

In order to distinguish these interactions from 
the SAMi-J family introduced above, which is
momentum-dependent (MD), in the following we will indicate them as
momentum independent (MI) interactions. Concerning the symmetry energy, 
several trends are considered, as shown in Fig.~\ref{eossym} (lower panel), 
leading to different values of the slope L, but close values of the symmetry energy at normal density  (J $\approx$ 30 MeV) (see also Table~\ref{sl})~\cite{coll2}. 
As we will show in the following, the possibility to probe several interactions
in the transport dynamics allows one to define the density regime explored
in low-energy nuclear reactions and to test the impact of density dependent
terms, such as the symmetry energy, on reaction observables.  


The ground state of the considered nuclei is determined by solving Eq.~(\ref{vlasov}) in the stationary limit. Since we work with test particles which
are usually associated with wave packets of finite
width, some surface contributions are already implicitly taken into account, both in the initialization and in the dynamics, in addition to 
the surface terms of the SAMI-J interactions. In our case, in
particular, we adopt triangular functions~\cite{Guar96}. 
Actually, we find that the optimal reproduction of the experimental features
(binding energy and charge radius) of selected nuclei is attained when 
suppressing the explicit surface terms, i.e.  $C_{surf} = D_{surf} = 0$.   
 Therefore we will adopt this choice in the following.

\begin{table*}[t]
\begin{center}
\begin{tabular}{|c|c|c|}
\hline
Effective interaction  & J [MeV] &  L [MeV]   \\
\hline
asy-soft & 30 & 14.8  \\
\hline
asy-stiff & 30.5 & 79  \\
\hline
asy-superstiff & 30.5 & 106  \\
\hline
SAMi-J27 & 27 & 29.9 \\
\hline
SAMi-J31 & 31 & 74.5 \\
\hline
SAMi-J35 & 35 & 115.2 \\
\hline
\end{tabular}
\caption{The values of the symmetry energy $J$ and its slope $L$ at normal density are reported for the Skyrme interactions adopted in our study.} 
\label{sl}
\end{center}
\end{table*}

\section{Dipole excitations in neutron-rich systems}
For the study of collective motion in nuclei, we neglect the collision integral in Eq.~(1).
Thus we are lead to consider 
the Vlasov equation, which represents the semi-classical limit of Time-Dependent Hartree-Fock (TDHF)
and, for small-oscillations, of the Random Phase Approximation (RPA) equations. In our calculations, 
a number of $1500$ t.p. per nucleon is considered, ensuring a good spanning of the phase space. 
{We will consider the following neutron-rich nuclei, spanning three 
mass regions:}
$^{68}$Ni (N/Z = 1.43), $^{132}$Sn (N/Z = 1.64),  $^{208}$Pb (N/Z = 1.54).

\subsection{Ground state properties}

\begin{table}[t]
{\renewcommand\arraystretch{1.2}
\begin{center}
\begin{tabular}{|c|c|c|c|c|}
\hline 
& \textbf{$\sqrt{\langle r^2 \rangle_n}$} [fm] &  \textbf{$\sqrt{\langle r^2 \rangle_p}$} [fm] &  \textbf{$\sqrt{\langle r^2 \rangle_n} -\sqrt{\langle r^2 \rangle_p}$} [fm] & \textbf{$\frac{B}{A}$} [MeV] \\
\hline
\multicolumn{5}{|c|}{\textbf{$^{68}$Ni}} \\
\hline
asy-stiff & 4.104 & 3.907 &  0.197 & 10.905 \\
\hline
SAMi-J31 & 4.102 & 3.898 & 0.204 & 9.050 \\
\hline
Exp & ---  & 3.857  & --- & 8.682  \\
\hline
\multicolumn{5}{|c|}{\textbf{$^{132}$Sn}} \\
\hline
asy-stiff & 5.062 & 4.781 & 0.281 & 10.365  \\
\hline
SAMi-J31 & 5.035 & 4.741 & 0.294 & 8.552  \\
\hline
 Exp & ---  & 4.709 & --- & 8.354  \\
\hline
\multicolumn{5}{|c|}{\textbf{$^{208}$Pb}}  \\
\hline
asy-stiff & 5.793 & 5.592 & 0.201 & 9.826 \\
\hline
SAMi-J31 & 5.735 & 5.536 & 0.199 & 8.042 \\
\hline
Exp & ---  & 5.501 & ---  & 7.867  \\
\hline
\end{tabular}
\end{center}}
\caption
{Neutron and proton root mean square radii, their difference, and binding energy for the three systems considered in our study, as obtained with asy-stiff (MI) and SAMi-J31 (MD) interaction. The experimental values, for charge radius and binding energy, are also indicated~\cite{dataref}.
} 
\label{tab:skin}
\end{table}

{
The numerical procedure that we adopt to define the ground state gives
charge radius and binding energy values which agree rather well with the
predictions of Hartree-Fock calculations~\cite{coll4} 
and allow to get a reasonable reproduction of experimental values~\cite{dataref}, as one observes in Table~\ref{tab:skin}, in the case of the MD parameterizations. 
MI calculations overestimate the binding energy, if one imposes to have
similar neutron and proton density profiles as obtained in the MD case
(see Table~\ref{tab:skin}).  
The values reported for the neutron skin thickness 
are in good agreement also with previous results obtained with the Sly4
Skyrme interaction~\cite{SarPRC2007}, though the latter predicts
smaller neutron and proton radii with respect to our results. 
On the other hand, a more diffuse neutron skin is observed in the case of 
the Relativistic Mean Field (RMF) calculations reported in 
Ref.~\cite{SarPRC2007} 
(see also Refs.~\cite{GaiPRC2011, GaiPRC2012} for further details). 
Isoscalar and isovector density profiles are shown in Fig.~\ref{radius},  
for the system  $^{132}$Sn and the three SAMi-J parameterizations 
adopted in our analysis. 
As evidenced in the left panel, a more diffuse density profile is obtained
when increasing the slope parameter L. From the inspection of the 
isovector density (right panel), it appears that this effect can be ascribed
to the development of a neutron skin.  Indeed a larger slope L 
(see {for instance} the SAMi-J35 parametrization)
is associated
with a steeper variation of the symmetry energy around normal density, thus
favoring the migration of the neutron excess towards the low-density 
nuclear surface. A similar behavior is seen for the $^{68}$Ni  and $^{208}$Pb ground state configuration and
also in the case of the MI interactions~\cite{coll2}. The trend observed for the dependence of the neutron skin thickness on the symmetry energy features is in agreement with previous investigations with other models~\cite{carPRC2010, piek12, GaiPRC2011, GaiPRC2012}.

\begin{figure}
\begin{center}
\includegraphics*[scale=0.36]{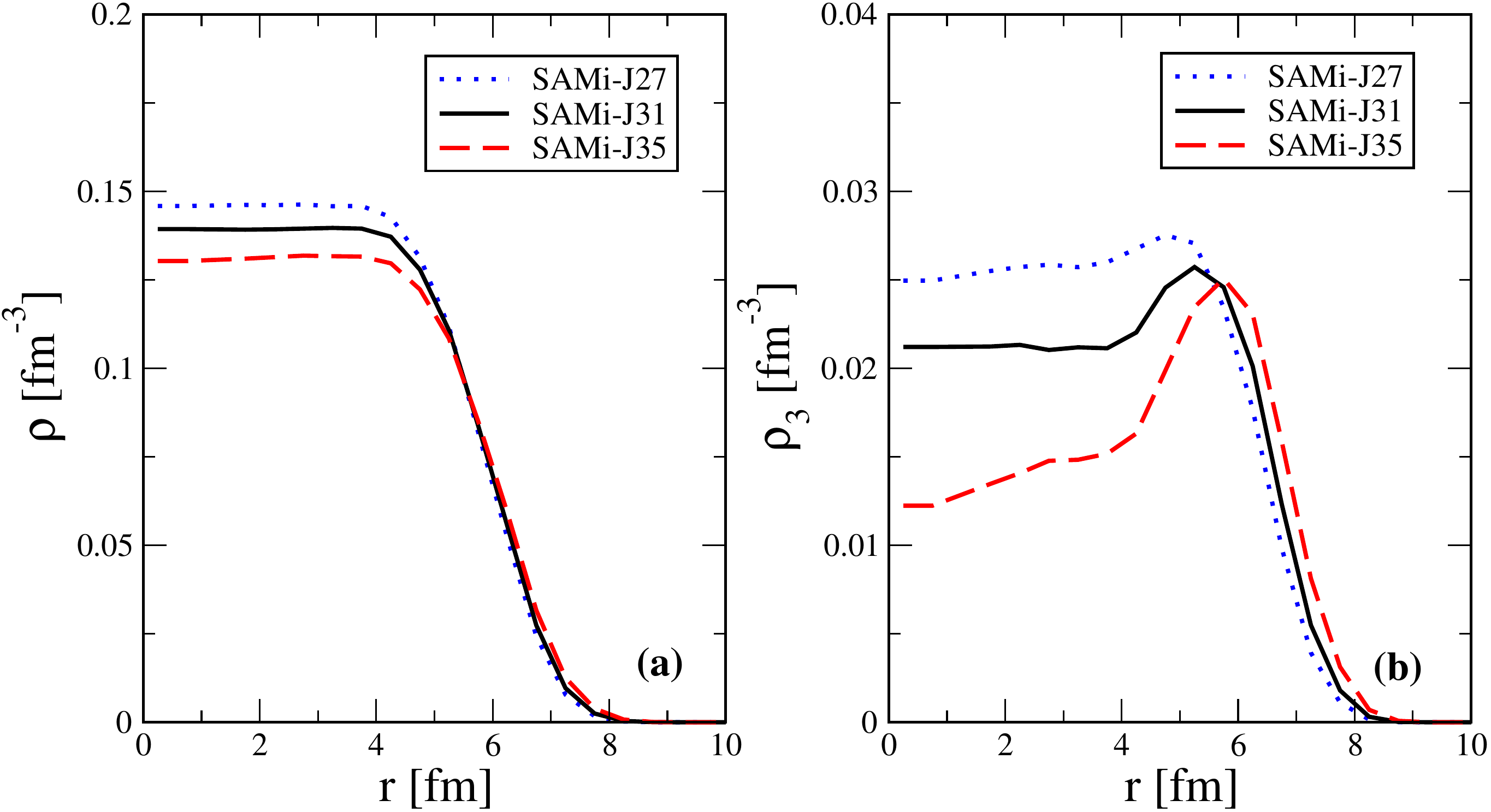}
\end{center}
\caption{(Color online) The isoscalar (left panel) and isovector (right panel) density profiles of $^{132}$Sn for the three SAMi-J parameterizations adopted in our study. Readapted from~\cite{zhePRC2016}.}
\label{radius}
\end{figure}

\subsection{Collective Dipole Response: isoscalar-isovector mixing}
{We concentrate our analysis on the E1 (isoscalar and isovector) response of nuclear systems. Thus we inject at the initial time the instantaneous
excitation} $\displaystyle V_{ext} =\eta_k \delta(t-t_0) \hat{D}_k$, at $t=t_0$, 
along the $z$ direction \cite{calAP1997,barPRC2012}, following the time evolution of the system
until $t=t_{max}$. 
Here $\hat{D}_k$ 
{indicates the operator inducing dipole excitations of isoscalar or isovector 
type ($k=$ S or V, respectively)}: 
\begin{align}
\hat{D}_S &= \sum_i \left ( r_i^2 - \frac{5}{3} \langle r^2\rangle \right) z_i; \\
\hat{D}_V &= \sum_i \left [ \tau_i \frac{N}{A} - \left (1-\tau_i \right) \frac{Z}{A} \right ]~z_i,
\label{dip_IV}
\end{align}
where {$\tau_i =0(1)$ for neutrons (protons) and  $\langle r^2\rangle$ refers to the mean square radius of the system under study.}
{It should be noticed that, in the general case of asymmetric
systems (with different N and Z numbers)}, the operator $\hat{D}_V$ also contains an isoscalar component.

{The strength function $S_k(E)$ is evaluated considering
the Fourier transform of 
{$ \displaystyle D_k(t)$, which is the expectation value of the time-dependent dipole moment:}}
\begin{equation}
 S_k(E) =\frac{Im(D_k(\omega))}{\pi \eta_k }~~,
\label{stre}
\end{equation}
where $\displaystyle D_k(\omega) =\int_{t_0}^{t_{max}} D_k(t) e^{i\omega t} dt$, 
with $E = \hbar\omega$.  

 {Introducing a gentle perturbation on the
ground state of the considered nucleus, we follow the time oscillations of the dipole moment, solving  Eq.~(\ref{vlasov}), until the final time
 $t_{max}=1800$~fm/c.}   
{A filtering procedure, as described in~\cite{reiPRE2006}, 
was applied in order to 
cure the problems connected to the finite calculation time. To this purpose, }
a smooth cut-off function was introduced such
that $D_k(t) \rightarrow D_k(t)\cos^{2}(\frac{\pi t}{2 t_{max}}) $.

{As discussed in Ref.\cite{zhePRC2016}, whereas in symmetric matter one
can isolate pure isoscalar and isovector excitations, in asymmetric systems
a mixing is generally observed, owing to the different amplitude of neutron
and proton oscillations. It is quite interesting to try to get a deeper 
insight into this effect and its dependence on the features of the effective
interaction employed.}

Fig.~\ref{isivsn132sami31} represents dipole oscillations and 
corresponding strength, as a function of the excitation energy $E$, as 
obtained for the system $^{132}$Sn and the SAMi-J31 interaction. 
The panels (a)-(d) correspond to an initial IS perturbation with  
$\eta_S = 0.5$~MeV~fm$^{-2}$, whereas an initial IV perturbation with 
$\eta_V = 25$~MeV has been considered in panels (e)-(h).

{The mixing between isoscalar and isovector excitations is rather evident.
Indeed the IS perturbation (panels (a)-(b)) also excites oscillations of the
IV dipole moment (panels ((c)-(d)).  
{In a similar way, when an IV perturbation is applied (panels (e)-(f)),
one also gets an isoscalar response (panels (g)-(h)).}}

{Let us start our discussion by looking at the features of the
isovector response (panel (f)).  Here we easily identify the IV GDR peak, 
 with $E_{GDR}\approx 14$ MeV. On the left, the low-energy region
(the so called PDR region) is moderately populated, with some strength
located between $E_1 = 9$ MeV and $E_2 = 11$ MeV.
Interesting enough, the contribution of the latter region is enhanced
when looking at the IS projection (panel (h)), where
the corresponding strength now aquires a similar amplitude as compared to the GDR. 
This observation already suggests that these low-energy modes are
mostly of isoscalar nature and is confirmed by the results obtained 
considering an initial IS perturbation (panels (a)-(b)). 
Indeed, in panel (b) one observes two important peaks with energies close to
$E_1$ and $E_2$, together with a moderate strength contribution in 
the IV GDR region ($E_{GDR} \approx 14$~MeV). 
 A considerable strength is located also in the high energy region of the spectrum ($E \approx 29$~MeV) and can be attributed to the IS GDR mode.  One can 
notice that also this mode has some mixed characted. 
In fact, a sizable (negative) contribution appears, at this energy, 
also in the IS projection corresponding to an initial IV perturbation (see panel (f)).  

From the results discussed above, one can conclude that in asymmetric systems
the normal modes are of quite mixed nature, so that they can be excited, 
though with different strength, by both IS and IV perturbations.
Thus it is appropriate to discuss essentially in terms of isoscalar-like
(i.e. mostly isoscalar) and isovector-like (i.e. mostly isovector) modes.
In particular, our analysis suggests that the modes located in the PDR
region are isoscalar-like; they contribute to the IV response because of their
mixed character~\cite{zhePRC2016}.
The dependence of these effects on the features of the nuclear effective
interaction is discussed in the next section.}

\begin{figure}
\begin{center}
\includegraphics*[width=0.45\textwidth]{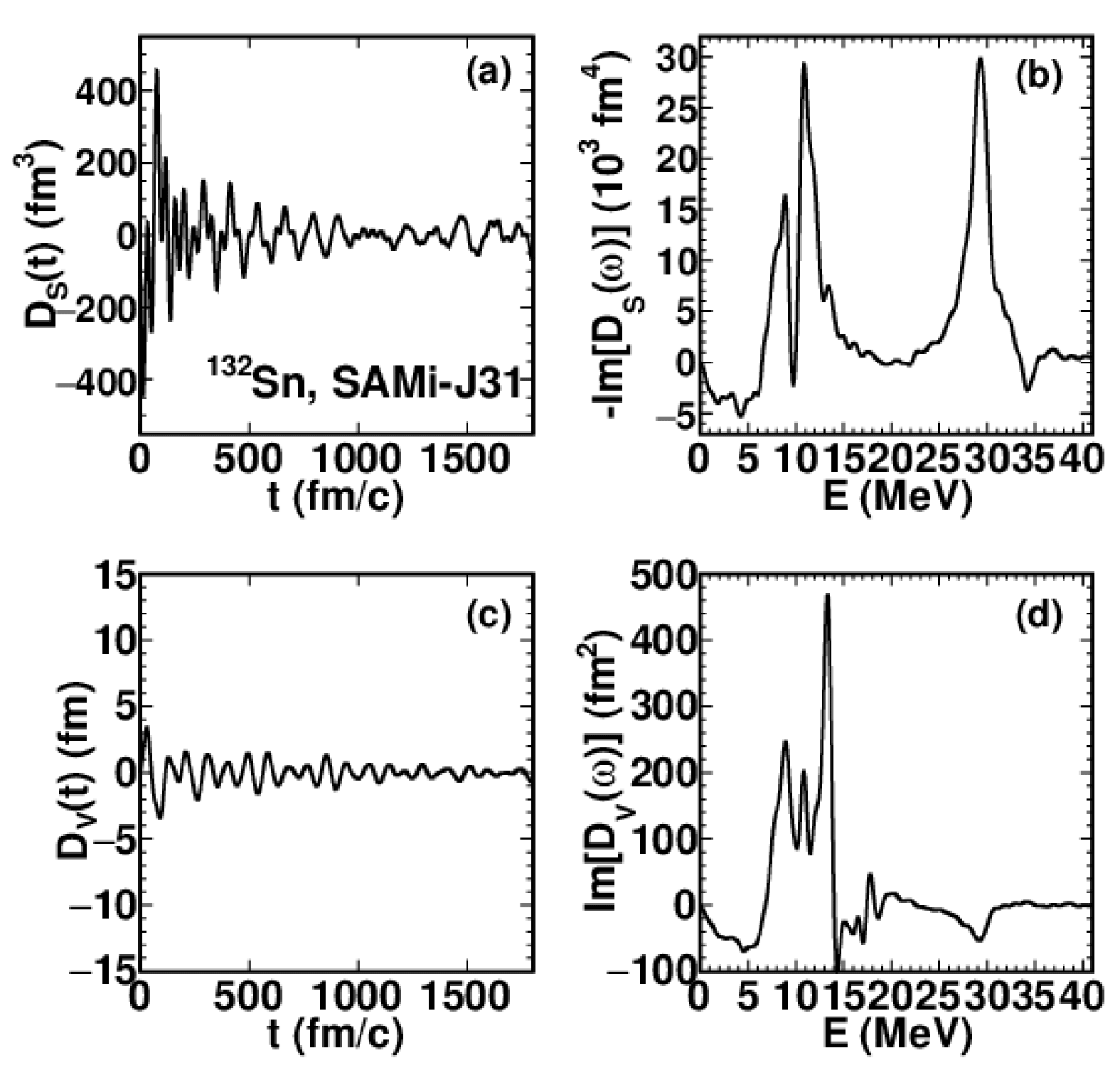}
\qquad 
\includegraphics*[width=0.45\textwidth]{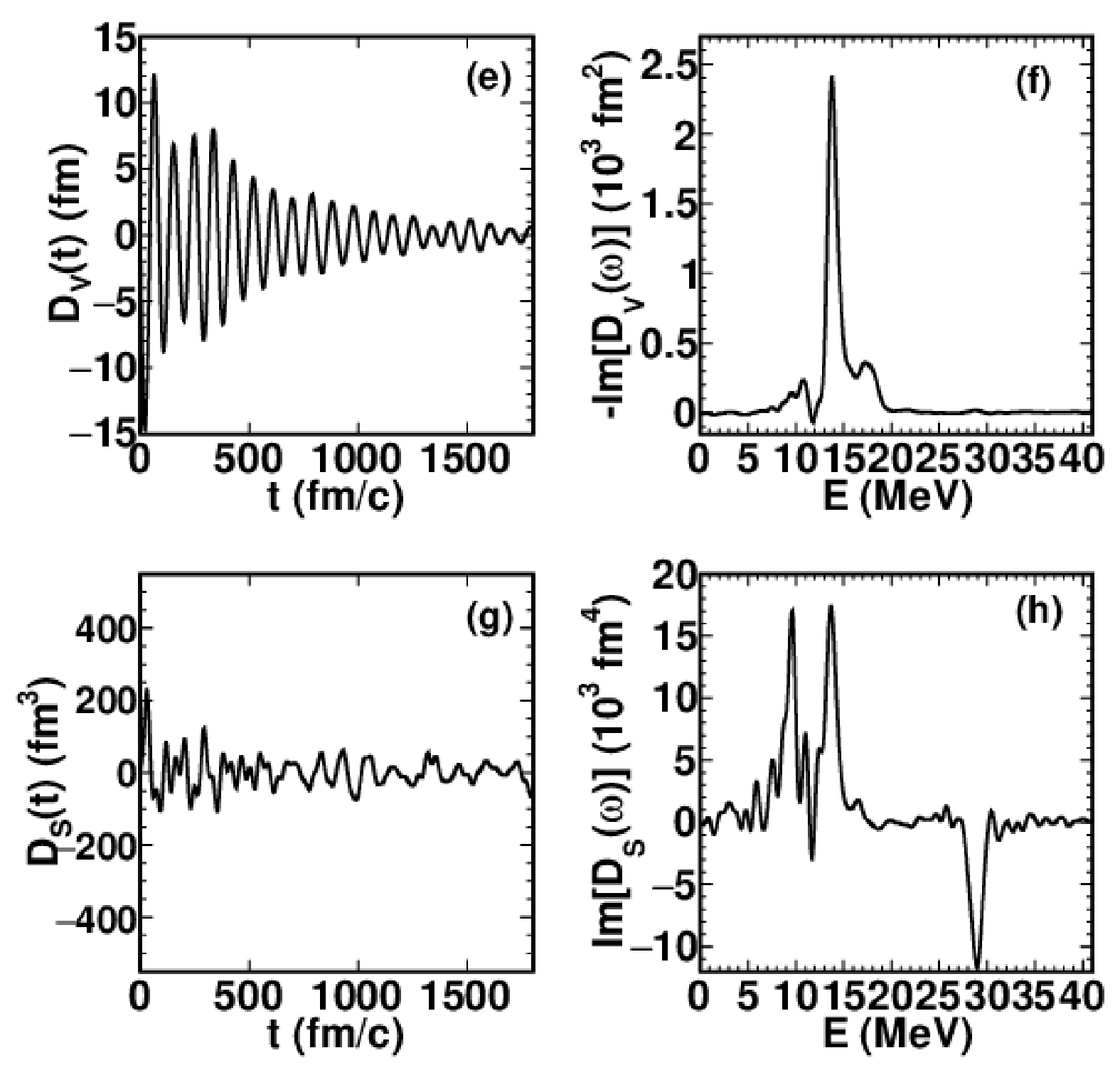}
\end{center}
\caption{The dipole oscillations and corresponding response functions for $^{132}$Sn and the SAMi-J31 interaction. Panels from (a) to (d) represent the results obtained with the initial IS perturbation and panels from (e) to (h) show the results obtained with the initial IV perturbation. Readapted from~\cite{zhePRC2016}.}
\label{isivsn132sami31}
\end{figure}
\subsection{Sensitivity to system size and effective interaction}
{Let us first discuss how the response of the system evolves
 in the three mass regions considered in 
this work}.  
{For the results shown in the following, we only consider the IS(IV)
response generated by a corresponding IS(IV) perturbation.}

\begin{figure}
\begin{center}
\includegraphics*[scale=0.36]{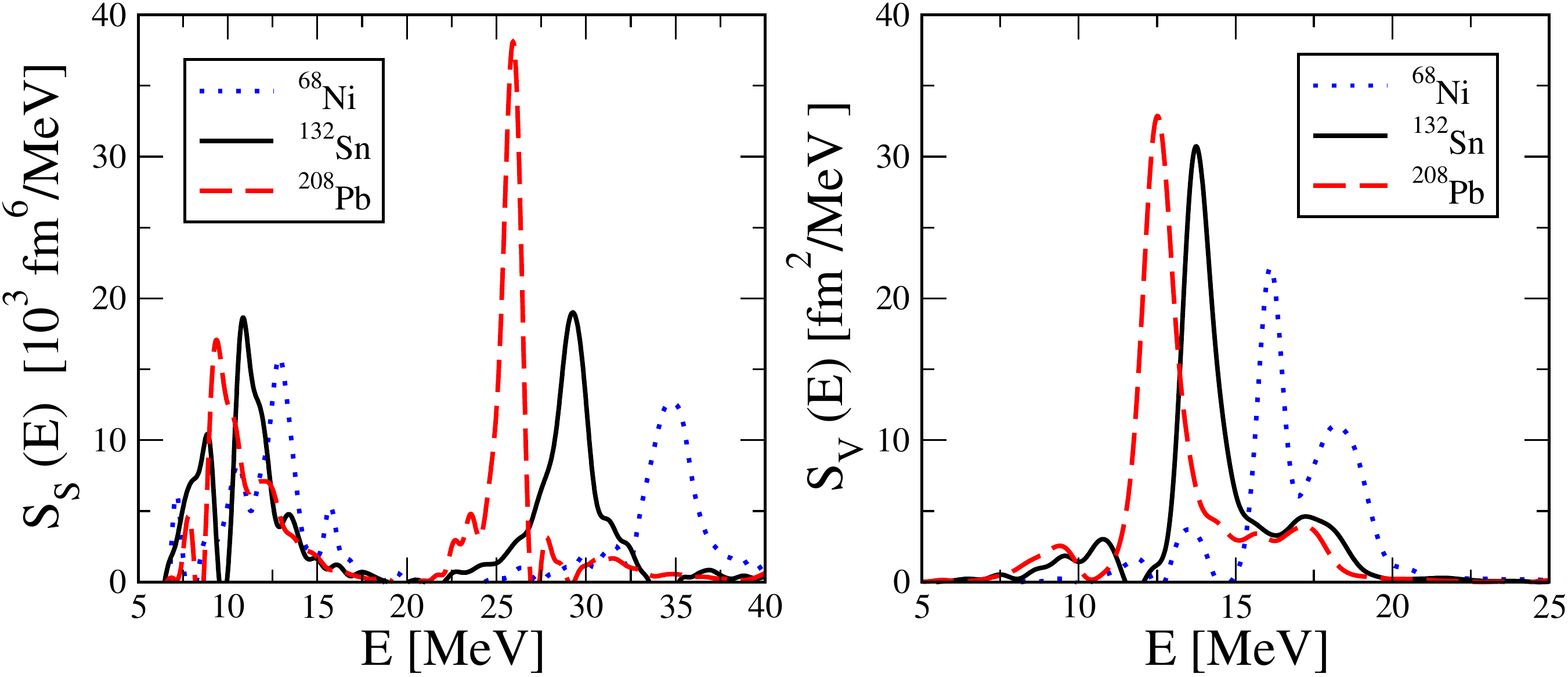}
\end{center}
\caption{(Color online) The strength function versus excitation energy for the three nuclei under study with SAMi-J31 interaction. The left panel refers to the IS strength, the right panel to the IV strength. The curves are normalized to the Energy Weighted Sum Rule (EWSR) of the IS (left panel) or IV (right panel) strength of the system $^{132}$Sn, respectively. Readapted from~\cite{zhePRC2016}.} 
\label{allmasses_isiv}
\end{figure}

In Fig.~\ref{allmasses_isiv} 
we show, 
for the SAMi-J31 effective interaction,
the strength function corresponding to the IS (left panel) and IV (right panel)
dipole response as a function of the excitation energy $E$. 
As a general remark, we observe that the response is shifted to lower energy
regions when increasing the system mass. 
Let us start discussing the 
IS response, whose spectrum is generally characterized by two main region of contributions: a large peak in the strength, which is associated with the compressional IS GDR mode and is located at high energy, above 25 MeV for all the nuclei under study, and a quite fragmented response, which is observed in the low-energy domain, in all cases below 15 MeV.  
The isoscalar-like nature of the isolated high-energy mode is considered well established, while its spreading width is still under investigation, although a
significant dependence on the size of the nucleus is already evidenced in Fig.~\ref{allmasses_isiv}. Concerning the low-lying energy modes, despite of the fragmentation, one can generally observe the emergence of 
two main peaks of comparable height with respect to the strength of the IS GDR, in agreement with previous results deduced within other semi-classical studies~\cite{urbPRC2012}, where these excitations have been preminently interpreted as surface modes. However, it is worth noting here that, owing to the coupling induced in neutron-rich systems and discussed above, these oscillations are then responsible also for the strength observed in the PDR region of the isovector response. This correspondence holds for the three nuclei considered, included the largest system, $^{208}$Pb, where the low-lying IS peaks tend to merge together. 
The features regarding the low-energy part of the dipole spectrum can be therefore addressed by looking also at the IV response. In this case, however, we observe, for all nuclei, that the IV projection of the PDR is quite smaller than the IV GDR (about one order of magnitude), in agreement with previous RPA calculations~\cite{mazPRC2012}. We conclude that the PDR region is mainly populated by 
isoscalar low-energy modes, which generally involve mostly nucleons belonging to the nuclear surface~\cite{zhePRC2016}. Thus the position and the relative importance of the different low-lying energy modes may reflect the shape (i.e. the 
volume/surface relative contributions)
of the density profile of the nucleus considered.

The reliability of our results is demonstrated by the good reproduction of 
the experimental data related to the IV GDR. 
Also the PDR region is reasonably reproduced, though a 
systematic overestimation is present in our calculations. 
This discrepancy might be probably attributed to the semi-classical treatment of surface effects. Indeed, these low-lying energy modes are mostly related to the oscillations of the most external nucleons. 
An improvement within the semi-classical framework can be probably achieved through a fine tuning of the coefficients $C_{\textup{surf}}$ and  $D_{\textup{surf}}$ in the Skyrme parameterizations.

\begin{figure}
\includegraphics*[width=0.3\textwidth]{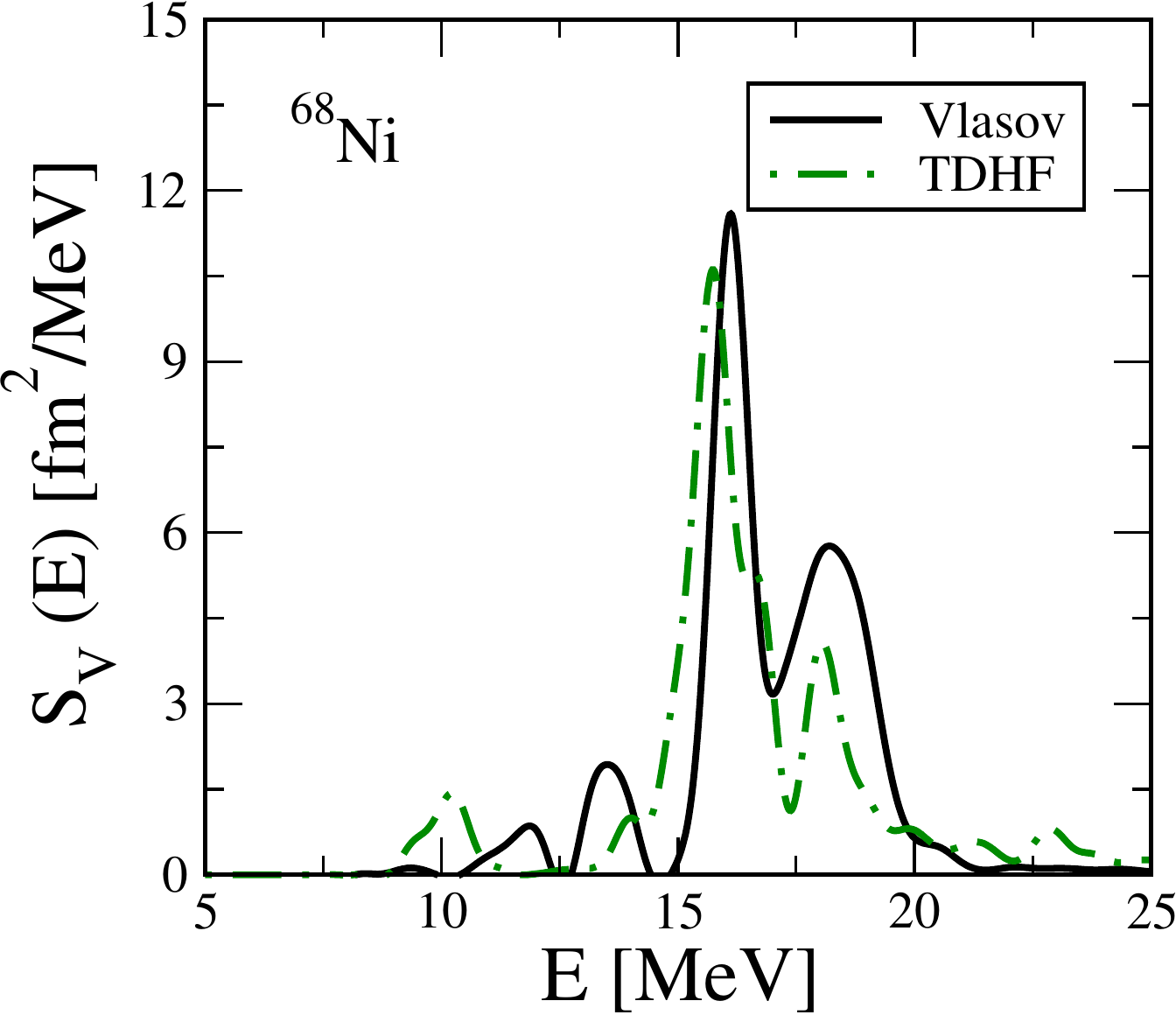} \quad \includegraphics*[width=0.3\textwidth]{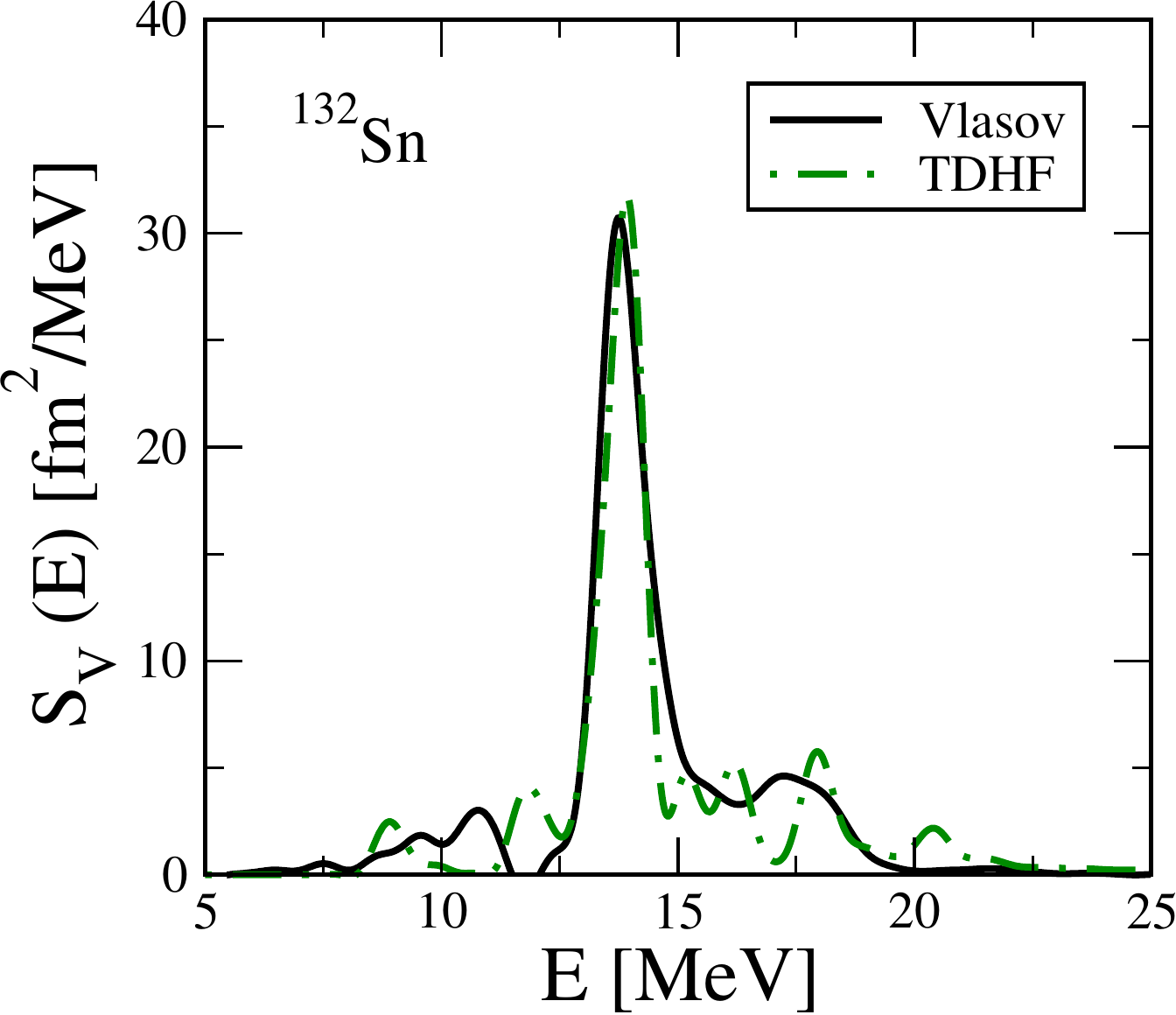} \quad \includegraphics*[width=0.3\textwidth]{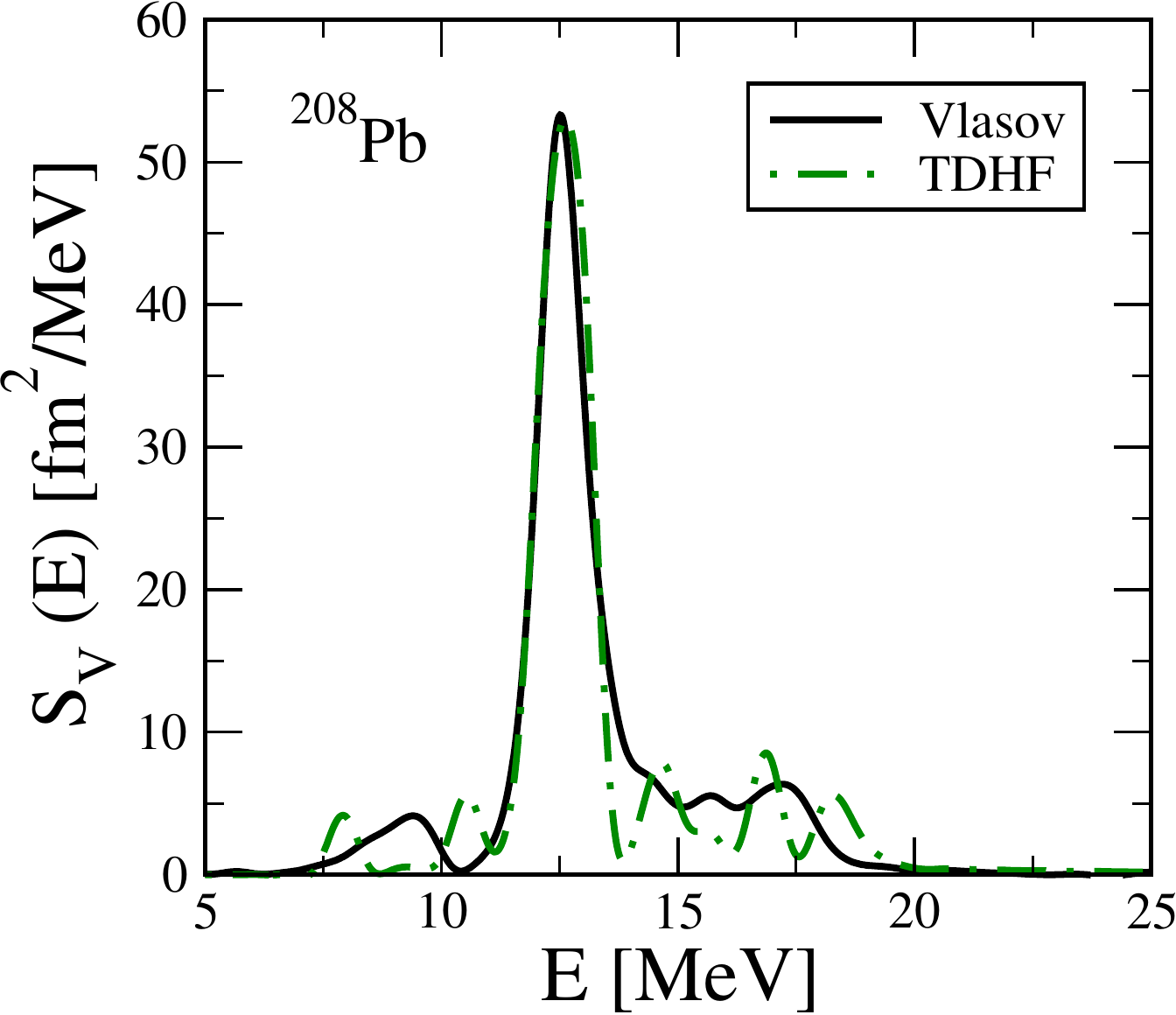}
\caption{(Color online) The IV strength function versus excitation energy for $^{68}$Ni (left panel), $^{132}$Sn (central panel) and $^{208}$Pb (right panel) with SAMi-J31 interaction, as obtained within the semi-classical Vlasov model or through a quantal TDHF calculation. Readapted from~\cite{zhePRC2016}.}
\label{sn132iv_vlasov_tdhf}
\end{figure}
To better explore this issue, in Fig.~\ref{sn132iv_vlasov_tdhf} we compare 
the IV dipole response extracted within our semi-classical Vlasov model 
to the results of standard TDHF calculations \cite{Denis}. 
Despite the general good agreement, 
especially for the heavier systems, of the main IV GDR peak energy resulting
from semi-classical and quantal approaches,   
significant differences between the two calculations are observed for 
the low-lying dipole modes. 
This comparison supports the conclusion that Vlasov results, in the PDR region, 
are affected by our numerical treatment of surface effects and by 
the lack of gradient terms of intrinsic quantal nature. 
A deeper investigation of the detailed structure of these excitations, both in semi-classical and quantal approaches, has thus to be envisaged.
{However, though the exact energy location of the PDR region is not
well reproduced, it is still worth to examine the dependence of the response
in this region on the effective interaction employed.} 

We will focus 
on the description of $^{132}$Sn.  
Let us look in particular at the change introduced in the spectrum when employing different SAMi-J parameterizations. We remind that this allows us to appreciate the sensitivity to the isovector channel of the interaction.  Qualitatively, looking at the central panel of Fig.~\ref{sn132isiv}, it appears that
the different peaks arising in the low-energy region  of the IV dipole response
become higher for larger $L$ values. 
{Moreover, the left panel indicates that also the IS response of the 
 lowest energy mode increases with $L$.
This is expected on the basis that a larger symmetry energy slope $L$ leads
to a larger coupling between isoscalar and isovector modes, as pointed out 
by calculations in asymmetric nuclear matter~\cite{barPR2005}.  Moreover, 
as seen in Fig.\ref{radius}, a stiffer symmetry energy leads to a ticker
neutron skin.  Thus surface modes become more important, with also a sizable
isovector component, owing to the neutron enrichmemt of the surface region. 
We conclude that the strength of the dipole response located in the PDR
region is quite sensitive to the symmetry energy parameterization and, in 
particular, to its slope $L$. 
{On the other hand, almost no sensitivity to the isovector channel
is seen for the energy position of the PDR strength}, as it is expected for IS-like excitations.  
}

\begin{figure}
\begin{center}
\includegraphics*[width=0.6\textwidth]{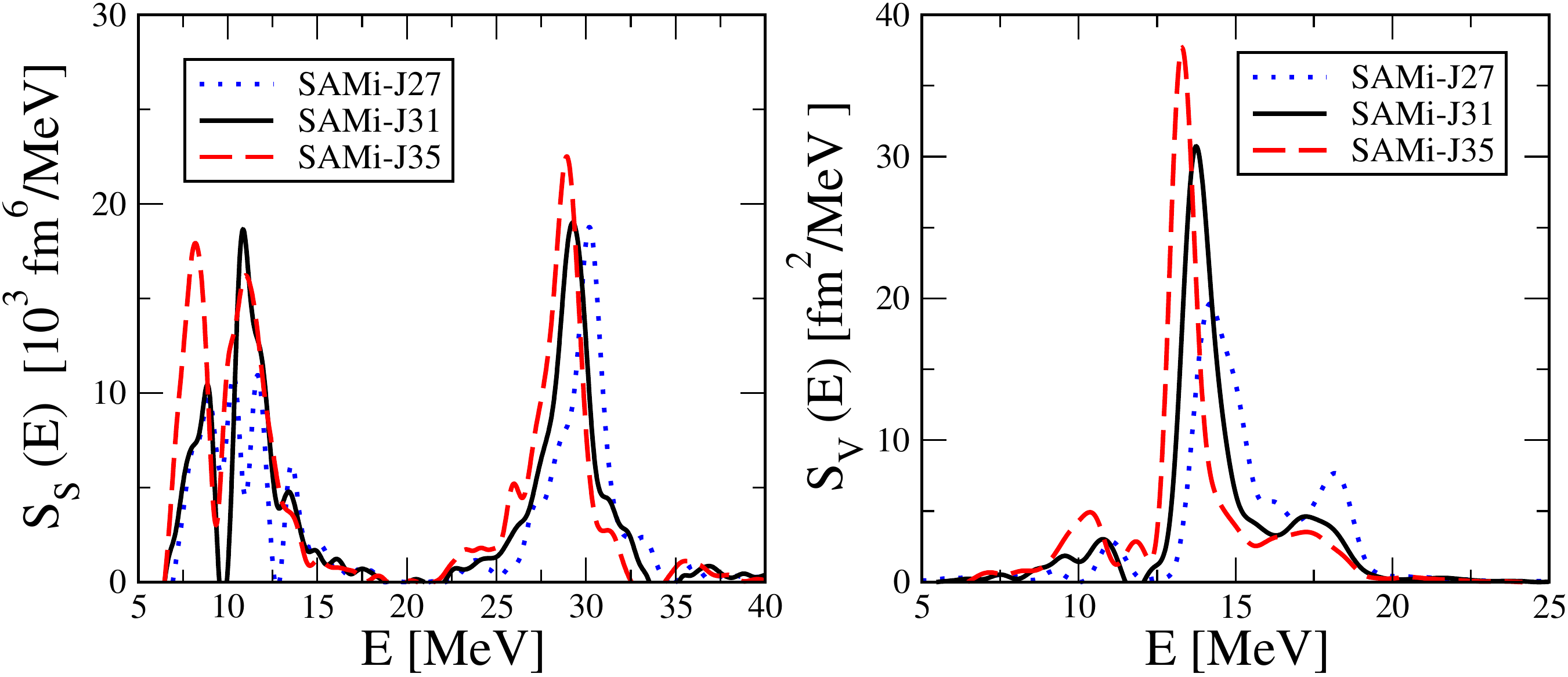} \quad
\includegraphics*[width=0.3\textwidth]{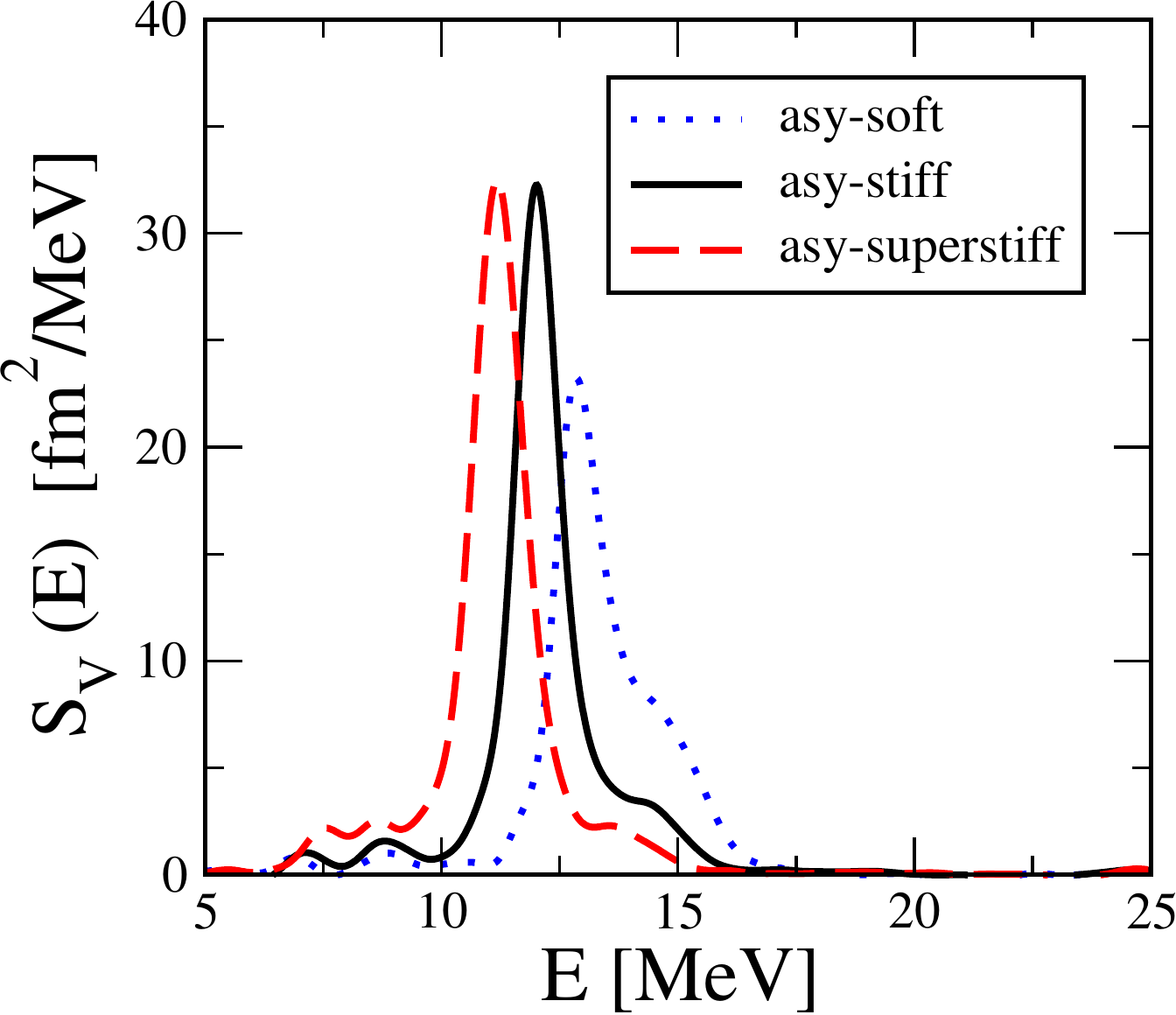}
\end{center}
\caption{(Color online) The IS (left panel) and IV (central panel) strength function versus excitation energy for $^{132}$Sn and the three considered SAMi-J interactions. Right panel: the same as for the central panel, but employing momentum independent interactions. Readapted from~\cite{zhePRC2016}.}
\label{sn132isiv}
\end{figure}
{
Other interesting features of the IV response can be discussed by taking 
into account also the results related to the MI Skyrme interactions, 
represented in the right panel of Fig.~\ref{sn132isiv}, still for $^{132}$Sn. 
As far as the energy of the IV GDR is concerned, one can see that it does
not evolve much in the SAMi-J case, whereas it shows a pronounced sensitivity
to the interaction in the MI case, being smaller for the $asy-superstiff$
parametrization.  
This suggests that the GDR energy reflects the value of 
the symmetry energy close to $\rho_c = 0.6~\rho_0$, which indeed can be taken as the average density of medium-heavy nuclei.  In fact, the three SAMi-J 
interactions have equal symmetry energy at $\rho_c$ (see Fig.~\ref{eossym}, panel (a)), whereas in the MI case (panel (b)) the symmetry energy is
smaller for the stiffer interaction.   
It is of particular interest to compare the results of asy-stiff and SAMi-J31 parameterizations, that show a close density behavior of the symmetry energy 
{(see Fig.~\ref{eossym})}.  In spite of this, 
one observes a higher frequency in the MD case. This can be ascribed to momentum
dependent effects, thus evidencing an interesting interplay between symmetry
energy and other terms of the effective interaction in shaping the features
of the nuclear response. 

Finally, a quite pronounced IV peak is observed 
{in the energy region above the GDR,} whose strength looks sensitive to 
the stiffness of the interaction.
As confirmed by the analysis of the transition densities, discussed below, 
this peak is associated with volume IV excitations.




\subsection{Transition densities}

{Additional information on the nature of the nuclear excitations, namely
on the mixing of IS and IV components, is gained through the study of the 
associated transition densities.  The latter describe how neutrons and protons
move in response to the external perturbation, thus helping to identify
the volume/surface character of the different modes~\cite{zhePRC2016}.  
}
{The transition densities, $\delta \rho_q$, essentially correspond to the density 
oscillations, around the ground state configuration, induced by the
initial perturbation. They can be calculated separetely for neutrons 
and protons. 
Exploiting the cylindrical symmetry of the system and 
making the same assumptions (linear response regime) as in  Ref.\cite{urbPRC2012}, one can write:  $\delta \rho_q(r,\cos\theta,t)=\delta\rho_q(r,t)\cos\theta$.
Thus, at each time step,  the transition densities can be finally 
extracted, by performing an angular average, just as a function of the radial coordinate $r$. }

\begin{figure}
\begin{center}
\includegraphics*[scale=0.36]{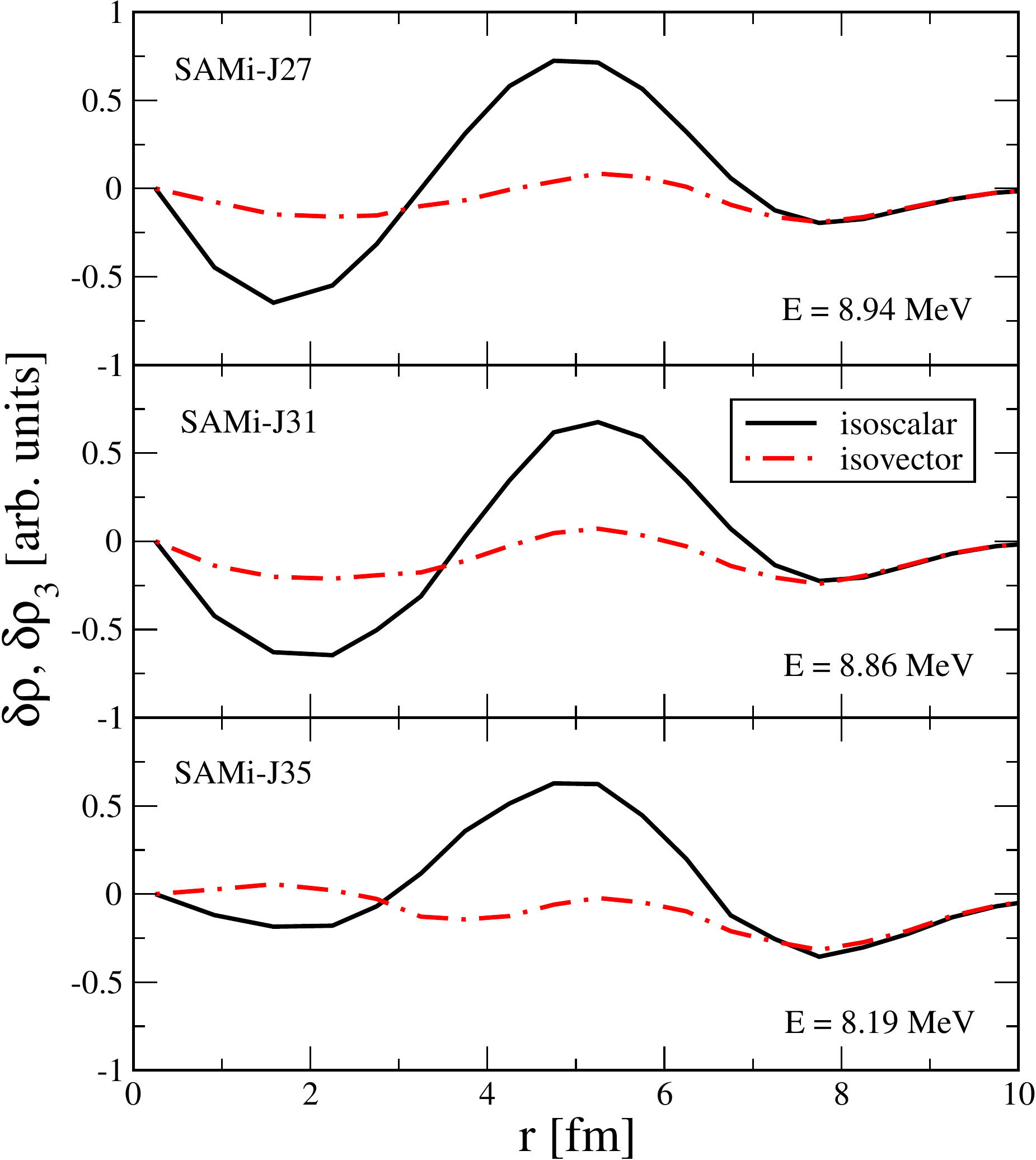}
\end{center}
\caption{(Color online) Radial dependence of the isoscalar and isovector transition density for the lowest energy dipole mode, for $^{132}$Sn and three SAMi-J interactions. Readapted from~\cite{zhePRC2016}.}
\label{Trhosn132samiIS}
\end{figure}

{As discussed above, different modes are excited by
 the delta function perturbation, $V_{ext}$, associated with the operator $\hat{D}_k$. Thus the observed transition densities will reflect the superposition
of the different oscillations.}
As explained in Ref.\cite{zhePRC2016}, 
in order to extract the contribution of a given mode, of energy $E$,  to the 
transition densities, one can 
consider the Fourier transform of $\delta{\rho}_q(r,t)$: 
\begin{equation} 
\delta{\rho}_q(r,E) \propto\int_{t_0}^\infty dt~  \delta{\rho}_q(r,t) \sin\frac{E t}{\hbar} . 
\end{equation}
{
Owing to the finite calculation time, the sine  function is multiplied by the same damping factor, as considered for the strength function $S_k(E)$. }

{
Since we are dealing with asymmetric systems, it is convenient to evaluate, for each mode,  
isoscalar and isovector transition densities (or, equivalently, neutron and proton
transition densities).  Then we expect the isoscalar component to be dominant in the case
of isoscalar-like excitation, where neutrons and protons oscillate in phase, though with 
different amplitude. On the other hand, isovector-like oscillations should be associated
with a dominant isovector component of the transition density.} 

In Fig.~\ref{Trhosn132samiIS}, we 
{display the isoscalar and isovector transition densities} associated with the lowest energy peak observed in the IS response 
(the PDR peak), for $^{132}$Sn and the three SAMi-J parameterizations adopted. 
As ahown in Fig.4, this excitation contributes also to the IV response. 
The curves in Fig.~\ref{Trhosn132samiIS} clearly manifest the isoscalar nature of the pygmy mode. Indeed the amplitude of the isoscalar density fluctuation is predominant overall, in contrast with the isovector one, which is generally rather small. However, a significant isovector density oscillation seems to involve the external part, where its contribution equals the isoscalar one. In fact, as a consequence of the neutron skin development, in this radial region only the neutron pratically oscillate and this fluctuation is responsible for the observed IV projection in the PDR region. Moreover, 
{the interactions characterized by a large slope $L$ lead to} 
an increase of 
both IS and IV transition densities (which practically coincide)
in the surface region. 
{Indeed, as discussed above, for increasing $L$, the system develops a thicker neutron
skin, thus the surface region becomes rather neutron rich and isovector effects are 
correspondingly enhanced in this region.
Hence, the analysis of the transition densities confirm the interpretation of the
strength observed in the PDR region discussed above.
}

\begin{figure}
\centering
\includegraphics*[width=0.48\textwidth]{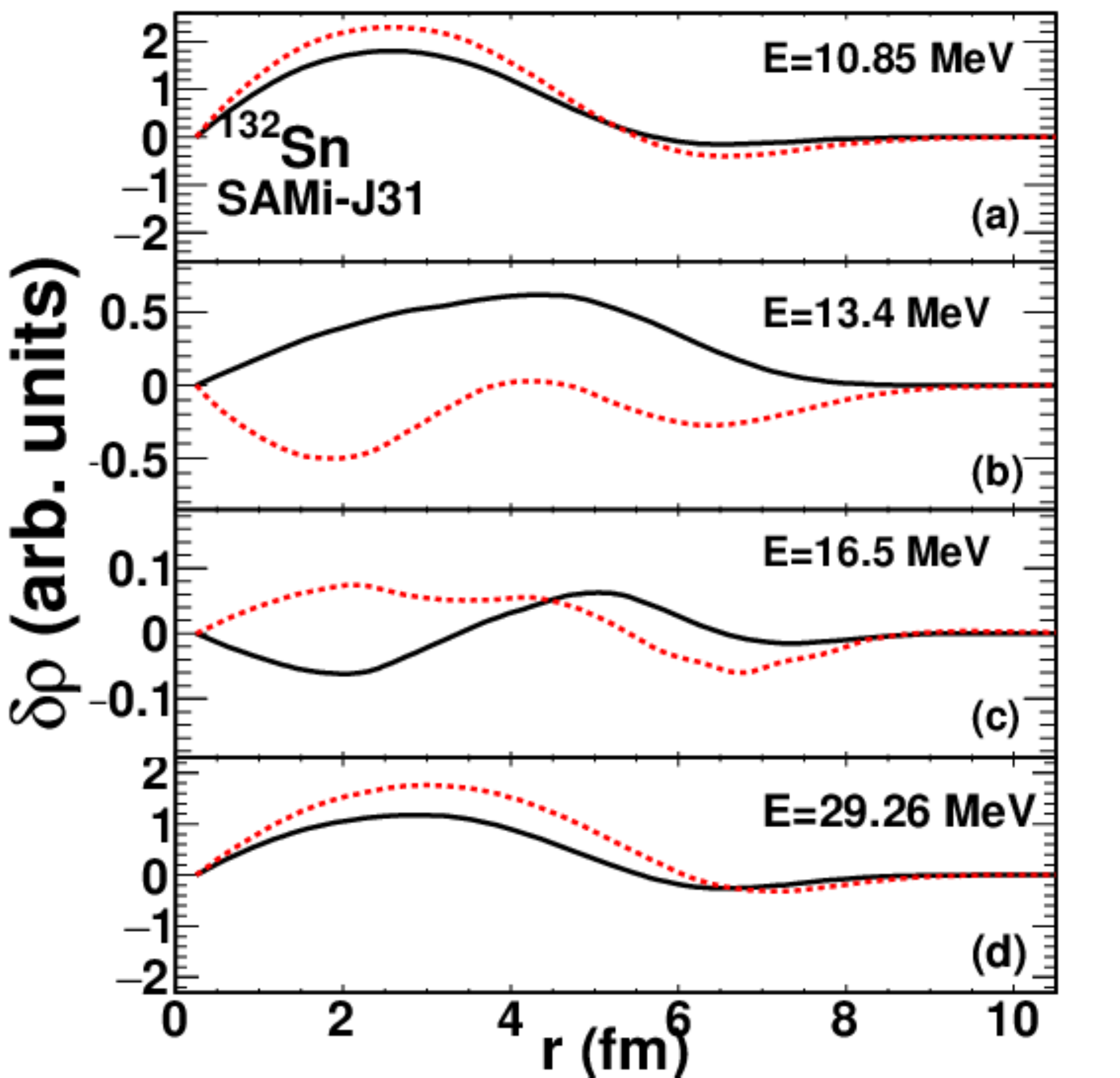} \quad 
\includegraphics*[width=0.48\textwidth]{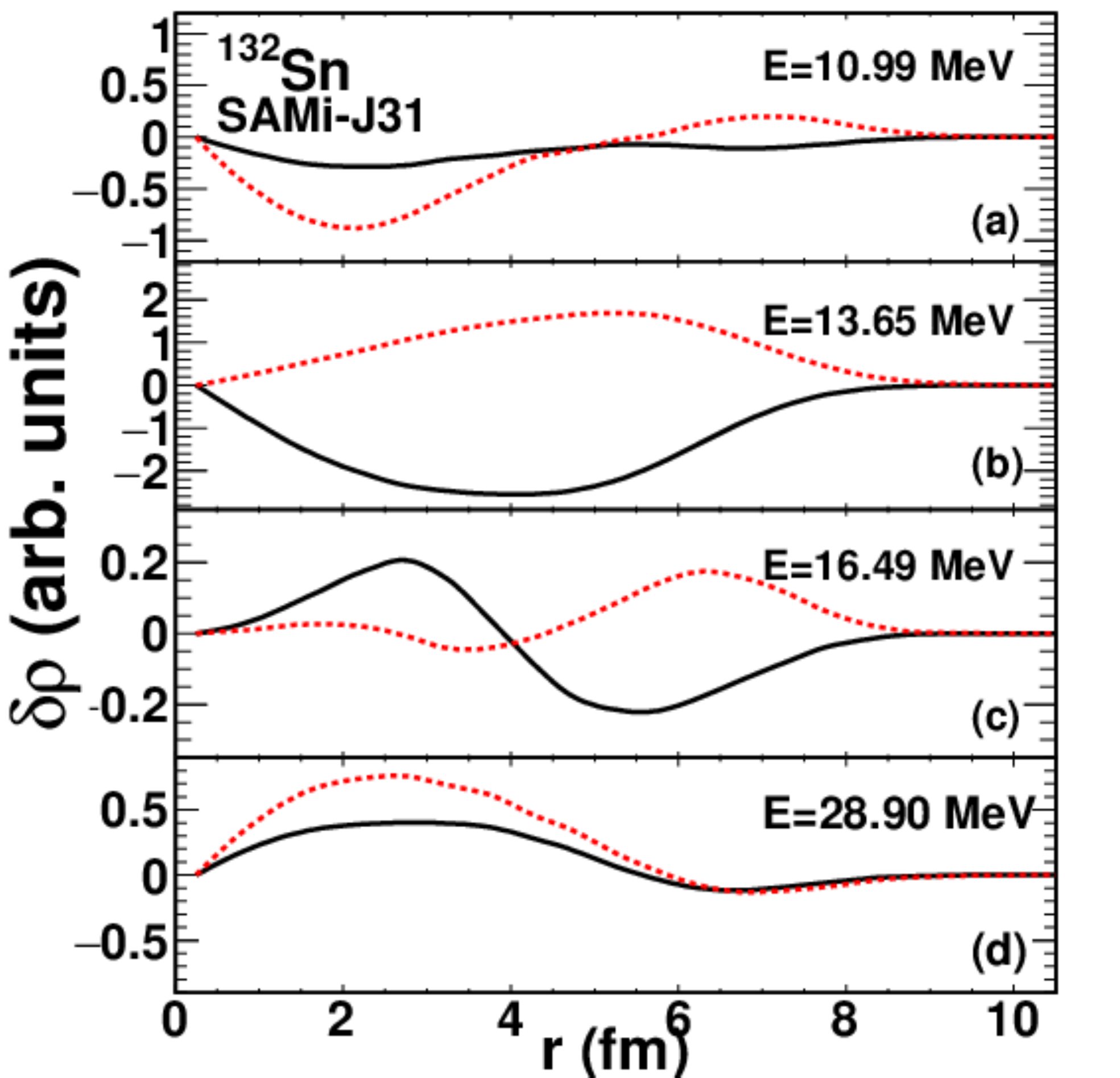}
\caption{(Color online) Left panels: the transition densities are displayed as a function of the radial coordinate $r$ for different excitation energies in $^{132}$Sn. An initial IS perturbation and the SAMi-J31 interaction are considered. Dashed lines refer to neutrons, full lines to protons. Right panels: similar to the left panels, but employing an initial IV perturbation. Readapted from~\cite{zhePRC2016}.}
\label{Trhosn132sami31IS}
\end{figure}
{
Fig.~\ref{Trhosn132sami31IS} shows the results relative to other modes 
which give a relevant contribution to the dipole strength, for the system $^{132}$Sn and  the SAMi-J31 interaction.  IS (IV) initial perturbations are considered in the left (right) panel.  
Let us comment first the features of the low-energy modes.  
As one can see in Fig.~\ref{Trhosn132sami31IS}, left panel (a), the transition densities
 of the second peak
identified in the IS response (around  $E_2 = 11$ MeV, see Fig.~\ref{sn132isiv}) indicate that
neutron and proton densities are in phase, though some mismatch is present.  
The mode has an isoscalar-like character, but with some mixing, leading to a contribution
also to the IV response. 

Indeed, when the same energy region is explored through IV excitations   
(Fig.~\ref{Trhosn132sami31IS}, right panel (a)), the difference between protons and neutrons becomes stronger. 
{We note that the feature described above, namely the splitting of 
the PDR strength into
an isoscalar contribution (at lower energy) and an isovector (more energetic)
component has been reported also  in recent theoretical and experimental analyses~\cite{Crespi2014,Edo1,Endre}. } 

Turning now to discuss the  highest energy isoscalar mode (Fig.~\ref{Trhosn132sami31IS}, panel(d)) we observe transition densities of considerable amplitude also in the internal part of the
system.  This confirms that this mode can be actually associated  with  the IS              dipole compression mode, of robust isoscalar nature.  By comparing left and right panels, one can
also notice that the features of this mode are practically not affected by the initial
perturbation. 

Concerning the modes that are essentially isovector-like (panels (b) and (c)), one can see
that also in this case the transition densities associated with IS or IV excitations 
are quite similar (compare left and right panels).
The transition densities nicely indicate that, whereas the standard GDR (panels (b)) is
essentially a surface mode, the higher energy mode (panels (c)) 
{deeply involves the nucleons belonging to the internal part, exhibiting a double oscillation, typical of Steinwedel-Jensen volume modes. } 
}

\section{Results for reaction dynamics}
Let us move now to discuss reaction dynamics between charge asymmetric systems, where charge 
equilibration takes place \cite{zhePLB}. 
{Full BNV calculations are performed for the reaction $^{132}$Sn+$^{58}$Ni at 10 MeV/A, considering several impact parameters leading to incomplete fusion and employing both MD and MI interactions. A proper number of t.p. (600 t.p./nucleon) is adopted to ensure a reasonable spanning of $f_q$ in phase space, as well as an acceptable computing time.}
Considering $^{132}$Sn as a projectile induces a sizeable charge asymmetry in the entrance channel, also allowing  
to explore possible reaction effects related to the neutron skin thickness, 
whose dependence on $L$ has been already addressed in the previous section~\cite{zhePRC2016} 
and shown in Fig.~\ref{radius}.  {Moreover, considering projectile (P) and target (T)
 with different N/Z ratios ($(\frac{N}{Z})_P=1.64$ and $(\frac{N}{Z})_T=1.07$ in the case
 considered), neutron and proton centers of mass do not coincide, thus creating an initial 
dipole moment which may trigger DD oscillations along the rotating reaction symmetry axis. 
Another interesting aspect of nuclear reactions at energies just above the Coulomb barrier 
is that pre-equilibrium nucleon emission starts to take place.   The two 
pre-equilibrium effects, 
namely nucleon and  $\gamma$-ray emission will be therefore addressed in the following~\cite{zhePLB}.   }

\subsection{The DD emission}
We first discuss dipole oscillations, 
following a collective bremsstrahlung analysis~\cite{baranPRL2001, parasPRC2016, baranPRC2009}. Similarly to Eq.~\ref{dip_IV}, the dipole moment is defined in coordinate space as:
\begin{equation}
D(t)=\frac{NZ}{A}(R_p-R_n),
\end{equation}
where $A=A_T+A_P$ is the total mass of the dinuclear system and $Z=Z_T+Z_P$ ($N=N_T+N_P$) is the proton (neutron) number. $R_n$ and $R_p$ denote the center of mass of neutrons and protons, respectively.
For the system considered, when the two nuclei touch each other, 
the dipole moment is equal to $D_i$ = 45.1 fm. 
The DD emission probability of photons with energy  $E_\gamma$ is given by ($E_\gamma=\hbar \omega$):
\begin{equation}
\frac{dP}{dE_\gamma} = \frac{2e^2}{3\pi \hbar c^3 E_\gamma} |D''(\omega)|^2, \label{gmdecay}
\end{equation}
where $D''(\omega)$ is the Fourier transform of the dipole acceleration $D''(t)$~\cite{baranPRL2001}. We will show the results for the average DD evolution, as obtained by considering 10 events for each specific calculation.   This allows one to avoid spurious oscillations caused by the numerical noise associated with the finite number of test particles.    We expect  that, according to both mean-field and two-body collisional effects, the DD oscillations will be damped. 
{Since microscopic calculations suggest that in-medium effects quench the 
two-body nucleon cross section~\cite{Mac94}, we also performed simulations multiplying the latter quantity by a global factor $f_{cs}$. Two cases (i.e. $f_{cs}$ = 0, corresponding to Vlasov calculations, and $f_{cs}$ = 0.5) are considered. 
Hence in the following we will explore, in addition to mean-field effects, also the influence of collisional damping and nucleon emission on the dipole oscillations.} } 
\begin{figure}
\begin{center}
\includegraphics*[scale=0.36]{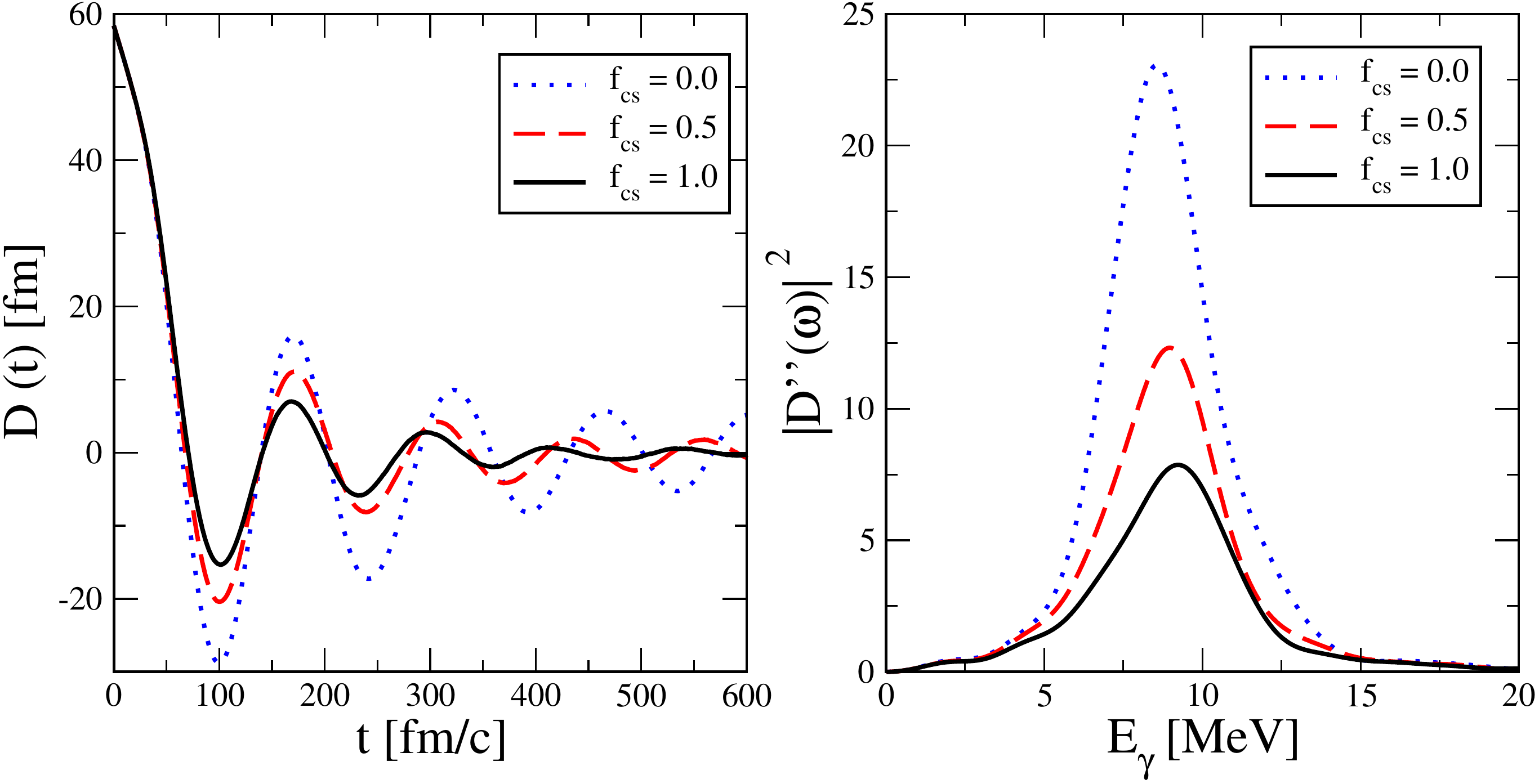}
\end{center}
\caption{(Color online) Left panel: The time evolution of DD for the SAMi-J31 EoS at b = 2 fm. Right panel: the corresponding power spectrum of the dipole acceleration. Results are plotted for different choices of the n-n cross section (see text). Readapted from~\cite{zhePLB}.}
\label{sn132ni58_dipole_spectra_fcs}
\end{figure} 
In Fig.~\ref{sn132ni58_dipole_spectra_fcs} (left panel), we plot the time evolution of the DD, at b = 2 fm, as obtained for the SAMi-J31 EoS and with the different choices of $f_{cs}$. {The initial dipole moment is quite large because at the initial time considered, the distance between the centers of mass of the two nuclei is 14 fm.} First of all, it can be interesting to compare the results related to the Vlasov case ($f_{cs}$ = 0) with the oscillations displayed in Fig.~\ref{isivsn132sami31}.  A considerable damping of the dipole oscillations is observed in Fig.~\ref{sn132ni58_dipole_spectra_fcs}, owing to possible non-linear effects and to nucleon emission, that cools down the system and reduces the initial charge asymmetry. One can also notice that the DD oscillations with the free n-n cross section  ($f_{cs}$ = 1) are damped faster than in the calculations associated with smaller $f_{cs}$ values. As a result, when neglecting the in-medium suppression of the n-n cross section, dipole oscillations are fully damped 
within about 600 fm/c. The corresponding power spectrum,  $|D''(\omega)|^2$, which enters the expression of the photon emission probability (see 
Eq.~(\ref{gmdecay})) is also represented in Fig.~\ref{sn132ni58_dipole_spectra_fcs} (right panel). {One can notice that the peak centroid is located at smaller energy with respect to the IV GDR in isolated nuclei. This is due to 
the elongated shape of the system at the initial reaction stage.  Moreover, one observes that the centroids of the power spectra do not depend too much on the cross section choice.  However,  a slight shift to lower energies is observed especially in the $f_{cs} = 0$ case, indicating that, in absence of two-body collisions,  the systems maintains the elongated shape for a longer time. Dissipation effects are clearly larger for the larger cross section ad the DD strength correspondingly decreases.  The calculations for the other impact parameters indicate that the DD signal is quenched in  more peripheral events, though similar features are observed with respect to the results for central collisions discussed above. }

\subsection{Sensitivity to the effective interaction}
Once the effect of the two-body collisional damping is clarified, we can look at the role of different ingredients of the nuclear effective interaction, in determining the energy spectrum, in analogy with the study of collective dipole modes carried out in  Section 3.3. The value of $f_{cs} = 1$ will be fixed therefore in the following, while all interactions introduced in Fig.~\ref{eossym} will be taken into account. The results are displayed in Fig.~\ref{sn132ni58_spectra_fcs1_orig_samiall}, for MI (left panel) and MD (right panel) interactions, respectively.

\begin{figure}
\begin{center}
\includegraphics*[scale=0.36]{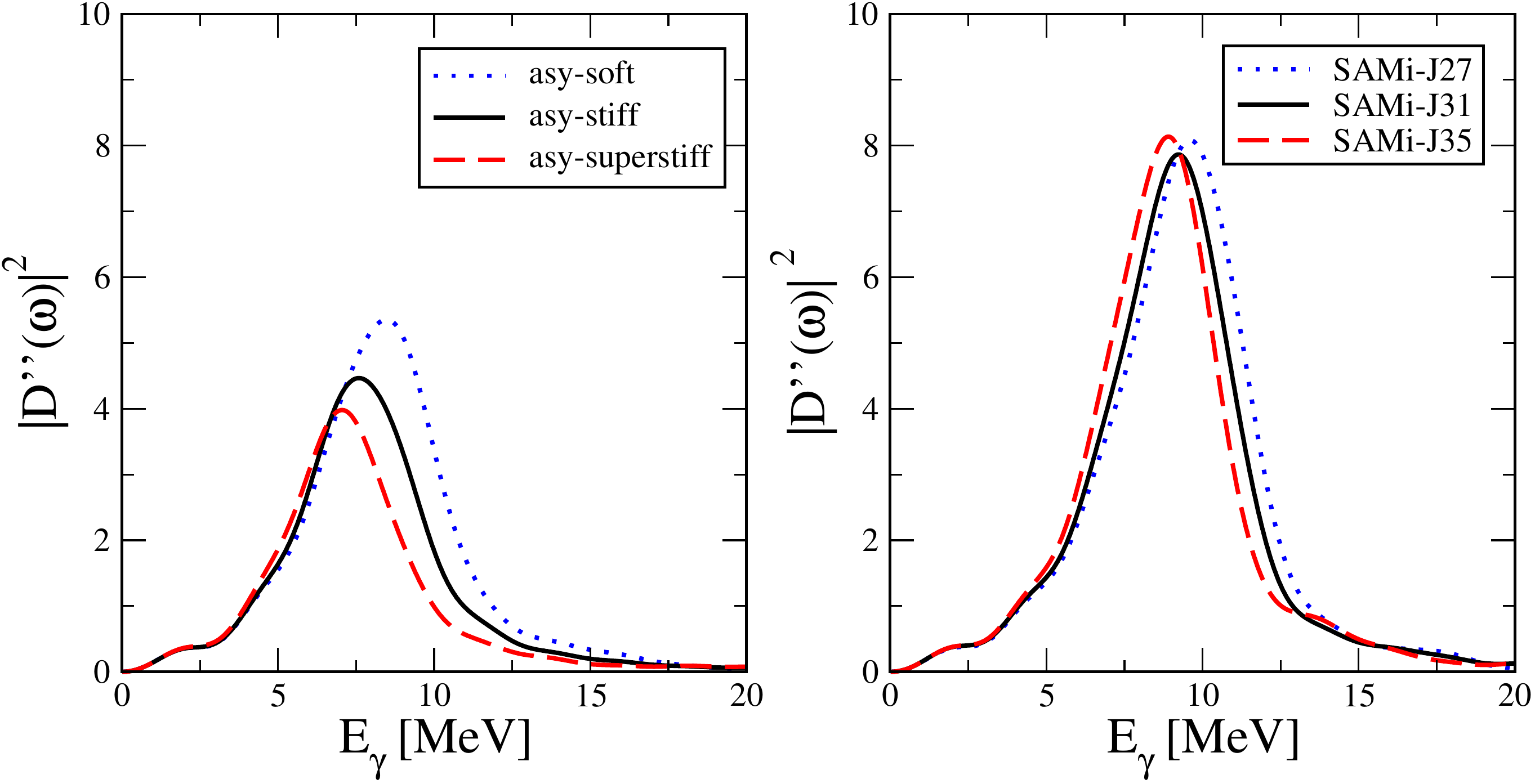}
\end{center}
\caption{(Color online) Power spectum of the dipole acceleration as obtained by employing MI (left panel) or MD (right panel) interaction, for $f_{cs} = 1$ and $b = 2$ fm (see text). Readapted from~\cite{zhePLB}.}
\label{sn132ni58_spectra_fcs1_orig_samiall}
\end{figure}
Several differences emerge between the power spectra obtained in the two cases {which are actually consistent with what we have already discussed in Section 3.3.   Indeed the peak centroid is insensitive to the effective interaction in the MD case, where a sizeable dependence is observed in the MI case. As already stressed above, the restoring force of IV dipole oscillations is essentially {ruled} by the symmetry energy. Thus we conclude that the also the DD features are determined by  the symmetry energy value in the density region around $\rho_c \approx 0.6~\rho_0$, the crossing point of the SAMi-J interactions. The lower frequency of the oscillations, with respect to the standard GDR, is due to the elongated shape of the system. In the MI case, the frequency of the power spectrum is higher for the $asy-soft$ case, in connection to to the larger value 
taken by the symmetry energy below normal density (see Fig.~\ref{eossym}(b)).  One can also observe that  a higher peak energy is also associated with a higher magnitude of the power spectrum, consistently  with previous studies~\cite{baranPRC2009}.  Moreover, the MD calculations are characterized by higher magnitude and frequency of the power spectrum, with respect to the MI results.  As already stressed in the case of the small amplitude excitations (see Fig.~\ref{sn132isiv}), this result 
is related to the influence of the effective mass on the features of collective {dipole} modes, for which it is well known that MD interactions 
yield a larger  
EWSR, which can better reproduce the experimental data~\cite{zhePRC2016}.
As discussed in Ref.\cite{zhePLB}, the results illustrated above 
can be grasped, in a rather schematic manner, in terms of the  damped oscillator model. Indicating by $\tau$ and $\omega_0$ damping time and oscillation frequency, respectively, the Fourier transform of the 
dipole acceleration can be written as~\cite{baranPRC2009}:
\begin{equation}
|D''(\omega)|^2 = \frac{(\omega_0^2+1/\tau^2)^2 D_i^2}{(\omega-\omega_0)^2 + 1/\tau^2}. \label{psana}
\end{equation}
From the above equation, it is clear that the DD emission is governed 
by the size of the initial dipole $D_i$, as expected, but it also reflects 
the amplitude of the oscillation frequency $\omega_0$. This is fully in line
with the results of Fig.~\ref{sn132ni58_spectra_fcs1_orig_samiall} (left): 
a larger energy centroid also corresponds to a larger strength.
One can also realize that a short damping rate, $1/\tau$, leading to strong
two-body collisional effects, quenches the strength  (see the denominator of Eq.~(\ref{psana})), as it is shown in Fig.~\ref{sn132ni58_dipole_spectra_fcs}. }   

\subsection{Nucleon emission}
Let us 
{discuss here the nucleon emission characterizing the pre-equilibrium stage}.
As we will show in the following, nucleon and light particle emission may carry out important information on
selected properties of the nuclear effective interaction. Indeed, for reactions
at Fermi energies, the isotopic content of such emission has already been shown to reflect the behavior of the
symmetry energy at subsaturation density. This is in fact the density region explored during the expansion phase of the nuclear composite system, when this emission mainly occurs~\cite{barPR2005,li08,zhangPLB2015}.
The nucleons emitted escape from the dense compact system, so that they can be traced back just looking at the particles belonging to
low-density regions ($\rho < 0.01~$fm$^{-3}$) at the final calculation 
time, $t_{max}$ = 600 fm/c.  
One observes that a larger nucleon emission is associated with MD interactions.  This can be attributed to the fact that the most energetic 
particles feel a less attractive mean-field potential when momentum-dependent
effects are turned on. 
Larger $f_{cs}$ values also lead to an enhanced pre-equilibrium emission, owing to the increased  n-n collision number.
On the contrary, the N/Z ratio decreases in the MD case and also in the calculations associated with a larger n-n cross section.  {Thus  it appears that,  
on the top to the expected sensitivity to the symmetry energy, 
the N/Z ratio 
also reflects some isoscalar features, such as effective mass and n-n collisions.  When the pre-equilibrium emission becomes more abundant,  the relative importance of isospin effects may become smaller and hence the N/Z ratio approaches 1.      This dependence on several aspects of the effective interation is seen also for the DD emission, as discussed above.}
However, one can pin down the sensitivity to the symmetry energy  
by inspecting in deeper detail the results corresponding to the three 
parameterizations considered in our study, either in the MI or MD case, with a given  $f_{cs}$ choice.
For instance, in the case of $f_{cs}$ = 1, the N/Z ratio obtained in
central collisions evolves from 2.049
to 1.774 when going from the asy-soft to the asy-superstiff symmetry energy parameterizations, whereas it changes from 1.433 to 1.687, in correspondence to the SAMi-J27
and SAMi-J35 interactions, respectively.
Bearing in mind that the N/Z ratio increases with the 
symmetry energy value (owing to the increased neutron repulsion), one can 
conclude that, in low-energy nuclear reactions, the pre-equilibrium emission
mainly tests the density region in between $\rho_c$ and $\rho_0$.
This statement is corroborated by the opposite trend of N/Z with respect to the slope, as obtained in the MI and MD cases.   Indeed
the crossing point of the symmetry energy (see Fig.~\ref{eossym})
is different for the two types of interactions, so that the ordering of 
the symmetry energy strength with L, in the density region considered, is opposite in the two cases. 
Therefore, combining the study of DD and pre-equilibrium nucleon emission, one has the possbility to probe the effective interaction in the low-density range $[\rho_c,\rho_0]$.

\subsection{Global analysis}
The results discussed above indicate that 
{the selected pre-equilibrium mechanisms, i.e. $\gamma$ radiation and nucleon emission,}  are influenced by 
several ingredients of the nuclear effective interaction. 
A deeper insight into this issue is got performing  a more global analysis, in order to explore mutual correlations between the features of the interaction 
and some proper observables~\cite{zhangPLB2015}.
The correlation coefficient $C_{XY}$ between the variable $X$ and observable $Y$ is 
usually defined as:
\begin{align}
C_{XY} &= \frac{cov(X, Y)}{s(X)s(Y)} \\
cov(X, Y) &= \frac{1}{n-1}\sum_{i=1}^n(X_i-{\bar X})(Y_i-{\bar Y}),
\end{align}
where $cov(X, Y)$ denotes the covariance,  $\bar A$ and $s(A)$ 
indicate average value and variance of A (=X or Y), respectively. These quantities are calculated from the considered set of MI and MD calculations, with different symmetry energy and n-n cross section parameterizations. 
It should be noticed that a linear correlation between  $X$ and $Y$
leads to $C_{XY}=\pm 1$, whereas, in absence of correlations, one gets 
$C_{XY}=0$.
Inspired by the results that have been presented so far,
we select three observables, which can be investigated experimentally~\cite{pierrPRC2009,zhangPLB2015}: the centroid ($E_{centr}$) and the integral of the DD 
emission strength, and the $N/Z$ ratio of the nucleons emitted. Three model parameters, which have been shown to impact significantly these observables, are considered: the symmetry energy slope $L$, the effective mass $m^*$ and 
the $f_{cs}$ value.
The correlations between the model parameters and the observables 
are presented in  Fig.~\ref{correlation} (see the solid bars) \cite{zhePLB}.
{ 
In this analysis, we intend to use the $L$ parameter to characterize the low-density behavior of the
symmetry energy. Thus the correlation functions have been evaluated excluding the SAMi-J27 and SAMi-J35 interactions. Indeed, within the SAMi-J family, the symmetry energy takes the
same value below normal density (at $\rho = \rho_c$, see Fig.1) and this feature could blur the sensitivity to $L$. 
One can see that an appreciable sensitivity to the slope $L$ can 
be identified for the DD centroid energy and for the N/Z of the pre-equilibrium emission.
A negative correlation is observed (denoted by the blue color); indeed both the centroid energy
and the pre-equilibrium N/Z decrease for stiffer (i.e. with larger $L$) interactions, as
discussed above. 
On the other hand, for the integral of the DD power spectrum the sensitivity to $L$ is 
overwhelmed by far by the huge dependence of the results on effective mass and n-n cross section.
For the sake of completeness, calculations have been performed also considering the full set of
MD and MI interaction (dashed bars in  Fig.~\ref{correlation}). As one can see, this
does not change the conclusions drawn above.
This analysis underlines and better quantifies the concurrent impact of several aspects
of the effective interaction on observables which, by construction, should be particularly
sensitive to the isovector terms, namely to the symmetry energy. 
In particular, it appear that it would be rather difficult to pin down the sensitivity to
the symmetry energy without fixing, at the same time, other ingredients, such as the effective mass
and the strength of the residual interaction. 
}

\begin{figure}
\begin{center}
\includegraphics*[scale=0.36]{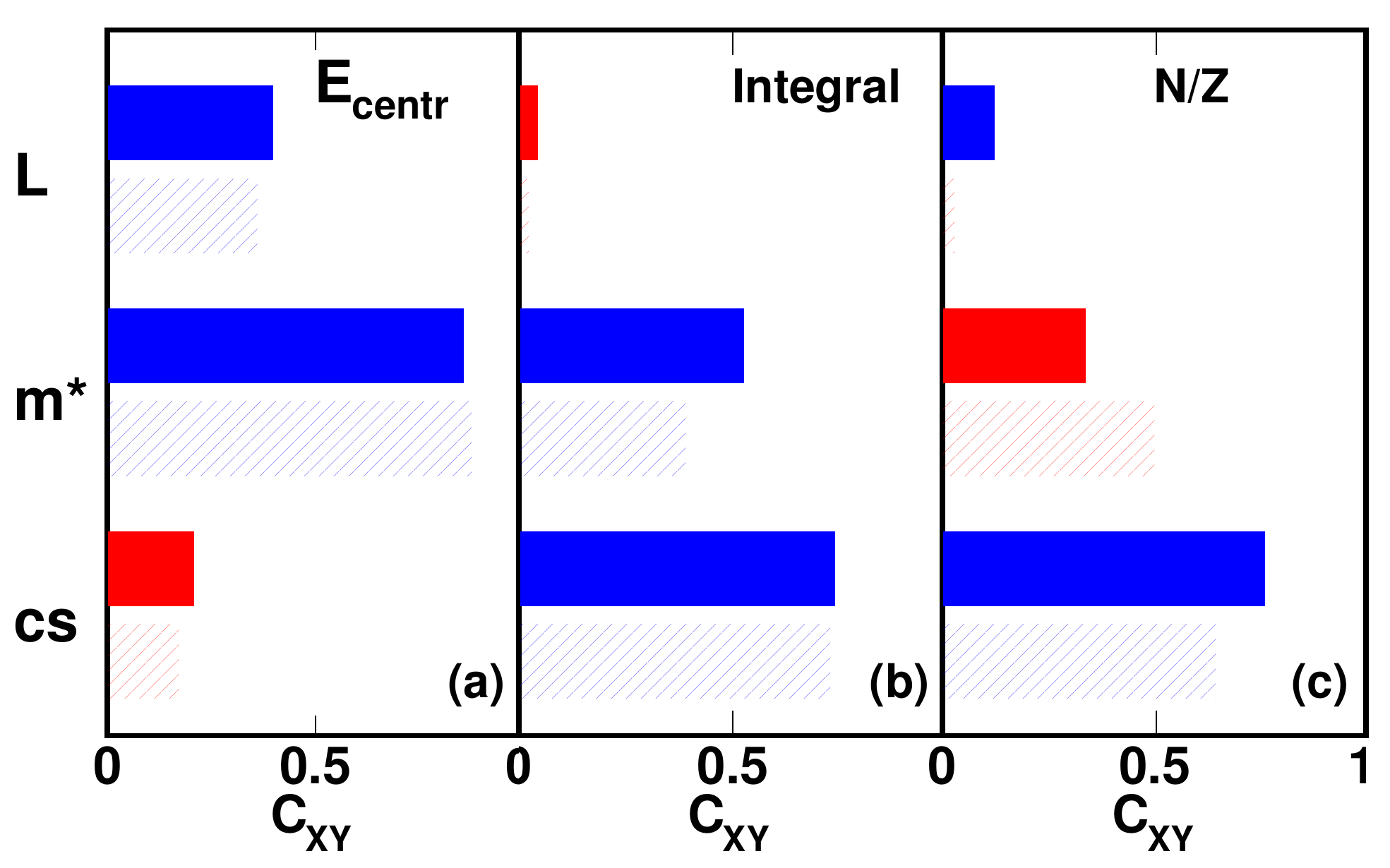}
\end{center}
\caption{(Color online) Negative (positive) correlations functions between model parameters and observables are indicated by blue (red) bars. Solid and shaded bars refer to different sets of calculations included in the analysis (see text for more details). Readapted from~\cite{zhePLB}.}
\label{correlation}
\end{figure} 

\section{Conclusions}
To summarize, in this work we have performed, within a semi-classical transport approach, a combined study of collective modes in neutron-rich nuclei 
and pre-equilibrium dipole radiation in low-energy nuclear reactions. Within the latter framework, pre-equilibrium nucleon emission has also been discussed. The aim of our investigation was indeed to assess the sensitivity of several observables involved in the reaction dynamics to the main ingredients of the nuclear effective interaction as well as to the in-medium n-n cross section. 

Concerning the analysis of collective excitations in neutron rich-systems, 
we aimed at getting a better insight into the features of the 
low-lying IV dipole response which is experimentally observed
in several neutron-rich nuclei~\cite{savPPNP2013}. 
Interesting features of the E1 nuclear response have been evidenced by exploring
three mass regions and considering a variety of effective interactions, 
mainly differing in the isovector channel.  
Inspecting both IS and IV response of the systems investigated, 
our analysis indicates the emergence, in neutron rich systems, of a strong mixing between isoscalar and isovector excitations, analogous to the one discussed also for infinite nuclear matter~\cite{barPR2005}, 
in agreement with previous semi-classical~\cite{urbPRC2012} or RPA~\cite{mazPRC2012} investigations. 
PDR excitations are mainly of isoscalar nature, however, because of mixing effects, 
some strength is observed also in the IV response, whose amplitude is rather sensitive
to the slope L of the symmetry energy around saturation. This observation is
associated with the appearance of a thicker neutron skin, in neutron-rich systems, for
increasing L values.  Hence our analysis confirms the important contribution of the 
study of low-lying dipole excitations to the symmetry energy debate.  
At last, it is also worth noticing that our results give a centroid energy of the IV GDR which is quite close to the experimental value as well as to RPA calculations~\cite{mazPRC2012}, also reflecting  
the value of the symmetry energy below normal density. 

Moreover, the results discussed here, in particular the link between the PDR strength, the neutron enrichment of the nuclear surface and the 
IS/IV mixing of the collective excitations, can be useful for the experimental search of the PDR and a for  
more accurate estimate of the corresponding strength. The latter, in turn, can 
provide information about the neutron skin thickness of the nucleus considered,
complementary to more direct measurements.  

As far as the study of pre-equilibrium dipole and nucleon emission is concerned, considering a variety of effective interactions in the calculations, we have asserted that these observables are also quite appropriate to explore 
the symmetry energy behavior below normal density, 
in the density range [0.6$\rho_0$, $\rho_0$]. However, though sensitive to the isovector channel of the interaction, we have brought out that these mechanisms strongly depend also on isoscalar terms, such as n-n correlations and 
momentum dependent terms.  A significant dependence on the latter terms, 
i.e. on the effective mass, is
observed also for the collective excitations 
investigated in Section 3~\cite{Jun}.  
Moreover, the sensitivity to several ingredients of the effective interaction has been recently pointed out also for the competition between fusion and quasifission processes in heavy ion reactions close to the Coulomb barrier~\cite{zhePRC2018}. 

Our analysis suggests that
a consistent study of collective excitations and low-energy reaction dynamics
may open the possibility, by a parallel investigation of different observables, 
to constrain at once the details of effective interaction and n-n 
cross section, together with  the low-density behavior of the symmetry energy.

\section*{Acknowledgements}
This project has received funding from the European Union's Horizon 2020 research and innovation programme under grant agreement N. 654002.

\end{document}